\documentclass[a4paper,fleqn,usenatbib]{mnras}
\usepackage{newtxtext,newtxmath}

\usepackage[T1]{fontenc}
\usepackage{ae,aecompl}

\usepackage{amsmath}	
\usepackage{graphicx}	

\usepackage{natbib}
\usepackage{longtable}
\usepackage{color}
\newcommand{\asec}{$^{\prime\prime}$}

\def\SigmaH2{$\Sigma $(${\rm H_2}$)}
\def\r1415{$^{14}$N/$^{15}$N}

\def\H{N$_{2}$H$^{+}$}
\def\D{N$_{2}$D$^{+}$}
\def\15N{$^{15}$NNH$^+$}
\def\N15{N$^{15}$NH$^+$}

\def\kms{\mbox{km~s$^{-1}$}}
\def\cmc{cm$^{-3}$}
\def\cmq{cm$^{-2}$}

\def\solm{\mbox{M$_\odot$}}

\def\Tex{\mbox{$T_{\rm ex}$}}

\def\kms{km\,s$^{-1}$}

\title[$^{14}$N/$^{15}$N in IRDC cores]{ALMA--IRDC II. First high-angular resolution measurements of the \r1415\ ratio in a large sample of infrared-dark cloud cores}

\author[F. Fontani]{F. Fontani$^{1,2}$,\thanks{E-mail: francesco.fontani@inaf.it}
           A.T. Barnes$^{3}$,
           P. Caselli$^{2}$,
           J.D. Henshaw$^{4}$,
           G. Cosentino$^{5}$,
           I. Jim\'enez-Serra$^{6}$,
            \newauthor
           J.C. Tan$^{5}$,
           J.E. Pineda$^{2}$
           and C.Y. Law$^{5}$
         \\
         $^{1}$INAF-Osservatorio Astrofisico di Arcetri, Largo E. Fermi 5, I-50125, Florence, Italy \\
$^{2}$Centre for Astrochemical Studies, Max-Planck-Institute for Extraterrestrial Physics, Giessenbachstrasse 1, 85748 Garching, Germany  \\
$^{3}$Argelander Institute for Astronomy, University of Bonn, Auf dem H\"{u}gel71, D-53121 Bonn, Germany \\
$^{4}$Max Planck Institute for Astronomy, K\"{o}nigstuhl 17, D-69117 Heidelberg,Germany \\
$^{5}$Space, Earth and Environment Department, Chalmers University of Technology, Chalmersplatsen 4, SE-41296 G\"{o}teborg, Sweden \\
$^{6}$Departamento de Astrof\'isica, Centro de Astrobiolog\'ia, E-28850 Torrej\'on de Ardoz, Madrid, Spain \\
         }

\date{Accepted 2021 March 5. Received 2021 March 5; in original form 2020 October 9}

\pubyear{2019}

\begin{document}
\label{firstpage}
\pagerange{\pageref{firstpage}--\pageref{lastpage}}
\maketitle

\begin{abstract}
The \r1415\ ratio in molecules exhibits a large variation in star-forming regions, especially when
measured from \H\ isotopologues. However, there are only a few studies performed at high-angular 
resolution. We present the first interferometric survey of the \r1415\ ratio in \H\ obtained with 
Atacama Large Millimeter Array observations towards four infrared-dark clouds harbouring 3~mm 
continuum cores associated with different physical properties. We detect \N15\ (1--0) in
$\sim 20-40\%$ of the cores, depending on the host cloud. The \r1415\ values measured
towards the millimeter continuum cores range from a minimum of $\sim 80$ up to a maximum of 
$\sim  400$. The spread of values is narrower than that found in any previous single-dish survey 
of high-mass star-forming regions, and than that obtained using the total power data only.
This suggests that the \r1415\ ratio is on average higher in the diffuse gaseous envelope of the cores,
and stresses the need for high-angular resolution maps to measure correctly the \r1415\ ratio 
in dense cores embedded in IRDCs.
The average \r1415\ ratio of $\sim 210$ is also lower than the interstellar value at the Galactocentric 
distance of the clouds ($\sim 300-330$), although the sensitivity 
of our observations does not allow us to unveil \r1415\ ratios higher than $\sim 400$. 
No clear trend is found between the \r1415\ ratio and the core physical properties.
We find only a tentative positive trend between \r1415\ and H$_2$ column density. However,
firmer conclusions can be drawn only with higher sensitivity measurements.
\end{abstract}


%
\section{Introduction}
\label{intro}

Nitrogen, the fifth most abundant element in the universe, has two stable isotopes: $^{14}$N 
and $^{15}$N. In the Solar System, the nitrogen isotopic fraction, \r1415, shows variations 
of an order of magnitude, from $\sim 50$ in carbonaceous chondrites (Bonal et al.~\citeyear{bonal10}) 
to $\sim 130-160$ in comets (Bockel\'ee-Morvan et al.~\citeyear{bockelee08},
Manfroid et al.~\citeyear{manfroid09}, Shinnaka et al.~\citeyear{shinnaka16}), up to $\sim 450$ in 
the atmosphere of Jupiter (Fouchet et al.~\citeyear{fouchet04}) and in the Solar wind 
($\sim 441$, Marty et al.~\citeyear{marty10}). The latter, in particular, is believed to 
represent the proto-Solar nebula (PSN) value (F\"{u}ri \& Marty~\citeyear{fem15}). 
This indicates an enrichment in $^{15}$N in pristine Solar System small bodies, but 
the causes of this enrichment, and in particular its relation with the chemical evolution 
of the PSN, are not yet understood.

In the past, a popular explanation for the $^{15}$N enrichment has been the isotopic
exchange reactions occurring at low temperatures (Terzieva \& Herbst~\citeyear{teh00}, 
Rodgers \& Charnley~\citeyear{rec08}), similar to those at the origin of molecular deuterium 
enrichment in cold dense cores (e.g.~Ceccarelli et al.~\citeyear{ceccarelli14} and 
references in there). 
This explanation has been challenged by theoretical works which have 
excluded low temperature reactions as the main way to enhance $^{15}$N
in molecular species (e.g.~Wirstr\"{o}m et al.~\citeyear{wirstroem12},
Roueff et al.~\citeyear{roueff15}, Wirstr\"{o}m \& Charnley~\citeyear{wec18},
Loison et al.~\citeyear{loison19}), and have also excluded significant nitrogen 
fractionation (both $^{15}$N enrichment and depletion) in the most abundant 
molecules during the chemical evolution of a star-forming core. Significant variations
in nitrogen fractionation are theoretically predicted only in extragalactic environments 
strongly affected by high fluxes of cosmic rays (Viti et al.~\citeyear{viti19}). 
However, these predictions are at odds with the large variations of \r1415\ measured in
star-forming cores. In fact, in low-mass pre-stellar cores the \r1415 ratio measured from \H\ varies 
from $\sim 330$ (Daniel et al.~\citeyear{daniel13},~\citeyear{daniel16}) to $\sim 1000$ 
(Bizzocchi et al.~\citeyear{bizzocchi13}, Redaelli et al.~\citeyear{redaelli18}, 
De Simone et al.~\citeyear{desimone18}, Furuya et al.~\citeyear{furuya18}), 
namely up to more than twice the PSN value, indicating a depletion rather than an 
enrichment of $^{15}$N. This depletion has been proposed to be due to the transition from 
atomic to molecular nitrogen in the earlier evolutionary stage of the cores by
Furuya \& Aikawa~(\citeyear{fea18}), although their models are not able to reach 
the antifractionation levels measured in pre-stellar cores. Some works
(Loison et al.~\citeyear{loison19}, Hily-Blant et al.~\citeyear{hily-blant20}) propose that only
a different dissociative recombination rate for \H\ and \N15\ could solve the problem 
and reconcile observations with chemical models, although it is hard to understand 
what could cause the different rate for the two species.
On the other hand, in CN and HCN the \r1415\ ratio seems more consistent with the PSN value (Hily-Blant et 
al.~\citeyear{hily-blant13a},~\citeyear{hily-blant13b}). In protoplanetary disks, the \r1415\ is 
found to be $\sim 83 - 156$ (Guzm\'an et al.~\citeyear{guzman17}) from observations of HCN 
isotopologues, while in TW Hya from CN is $\sim 323$ (Hily-Blant et al.~\citeyear{hily-blant17}),
i.e. in between the PSN and the cometary values. Overall, these findings suggest a high variability 
of the \r1415\ ratio, which depends on both the evolutionary stage of the source and the 
molecule observed.

Besides low-mass star-forming cores, a growing observational effort has been devoted to the 
study of the \r1415\ ratio in high-mass star-forming regions, based on the evidence that the Sun 
was likely born in a crowded stellar cluster including stars more massive than $\sim 8$\solm\ 
(e.g. Adams~\citeyear{adams10}, Pfalzner et al.~\citeyear{pfalzner15}, Lichtenberg et al.~\citeyear{lichtenberg19}). 
Traces of the interaction between these stars and the primordial Solar System are recorded in 
meteoritic material, where anomalous high abundances of daughter species of Short-Lived 
Radionucleides (SLRs), produced by nearby high-mass stars and ejected during the early 
evolution of the Solar System, are measured (e.g.~Portegies Zwart et al.~\citeyear{portegies-zwart18}).
Therefore, the birthplaces of massive stars and clusters are ideal targets to investigate the relation 
between the $^{15}$N enrichment in pristine Solar System material and its birth environment.

Even in this case, single-dish surveys of massive star-forming regions 
indicate large variations in the \r1415\ ratio. The highest variability is found, again, in \H\ 
($\sim 180-1300$, Fontani et al.~\citeyear{fontani15}), while values distributed in between the cometary 
values and the PSN one are found in CN, HCN, and HNC (Adande \& Ziurys~\citeyear{aez12},
Fontani et al.~\citeyear{fontani15}, Zeng et al.~\citeyear{zeng17}, Colzi et al.~\citeyear{colzi18a}, 
\citeyear{colzi18b}). The previous studies also suggest that the evolution does not seem to play a role, 
in agreement with the predictions of models (e.g.~Roueff et al.~\citeyear{roueff15}), although they 
cannot predict the broad measured range of isotopic ratios. 
However, the observations mentioned above were obtained with single-dish telescopes, thus providing 
average values of the \r1415\ ratio on linear scales of $\sim 0.1 - 1$~pc, which can contain gas with 
different conditions, due to the inner physical and chemical complexity
of high-mass star-forming cores. Hence, these observations could miss the presence of spots 
enriched (or depleted) in $^{15}$N on linear scales $\leq 0.1$~pc, i.e. smaller than the telescope beam. 

The few follow up observations performed at high-angular resolution have indeed suggested that local 
gradients in the \r1415\ ratio can be found at smaller linear scales: Colzi et al.~(\citeyear{colzi19}) have
observed the high-mass protocluster IRAS 05358+3543 with a linear resolution of $\sim 0.05$~pc, 
or $\sim 10000$ au, and found that the \r1415\ ratio in \H\ shows an enhancement of a factor $\sim 2$ (from $\sim 100 - 220$ 
to $\geq 200$) going from the inner dense core region to the diffuse, pc-scale, envelope, and interpreted 
these results as the consequence of selective photodissociation (Heays et al.~\citeyear{heays14}, Lee et al.~\citeyear{lee21}).
The latter mechanism, due to self-shielding of $^{14}$N$_2$, for \H\ predicts a decrease of the \r1415\ ratio 
in regions exposed to external UV irradiation, as indeed found by Colzi et al.~(\citeyear{colzi19}).
On the other hand, in embedded, not irradiated regions, the \r1415\ ratio in \H\ should not change, and 
this has been recently confirmed by observations with linear resolution of $\sim 600$ au of the 
protocluster OMC--2 FIR4, which have revealed a constant \r1415\ in the embedded protocluster
cores (Fontani et al.~\citeyear{fontani20}). However, these are, at present, the only two studies
performed at high-angular resolution in high-mass star-forming regions, and the results 
need to be corroborated by similar studies both towards other sources and in different molecules.

When investigating the birthplace of massive stars, infrared-dark clouds (IRDCs) are certainly ideal
targets. IRDCs are cold ($T\leq 25$ K, Pillai et al.~\citeyear{pillai06}), dense ($n{\rm (H_2)} \geq 10^{4-5}$ cm$^{-3}$) 
and highly-extinguished ($A_{\rm v} \geq 100$ mag and $N{\rm (H_2)} \geq 10^{22}$ \cmq, Butler \& Tan~\citeyear{bet09},
~\citeyear{bet12}) molecular clouds, first observed in extinction against the bright mid-IR Galactic background 
(Perault et al.~\citeyear{perault96}, Egan et al.~\citeyear{egan98}). 
Despite their importance for star-formation, they have been poorly investigated so far in nitrogen fractionation.
Zeng et al.~(\citeyear{zeng17}) measured with the IRAM-30m telescope the \r1415\ towards cores 
belonging to four IRDCs from HCN and HNC, and found once more a large variability (\r1415\ $\sim 70 - 800$ 
in HCN and $\sim 161 - 541$ in HNC), with the lowest values belonging to the lowest density regions. 
But, again, these are average values over $\sim 0.5-1$~pc linear scales that encompass gas with different 
physical and chemical properties. Moreover, these measurements also depend on the $^{13}$C/$^{12}$C 
ratio, which has been found to be affected by physical conditions as well as time evolution (Colzi et al.~\citeyear{colzi20}).

In this paper, we present the first interferometric study of the \r1415\ ratio towards a large sample of 
IRDC cores in \H\ at high-angular resolution. The \r1415\ ratio is derived from observations of \H\ 
and \N15\ (1--0) towards the dense cores of four clouds in which the physical and chemical structure 
has already been extensively studied at various wavelengths, from the mid-infrared to the sub-millimeter
(Butler \& Tan~\citeyear{bet09},~\citeyear{bet12}, Hernandez et al.~\citeyear{hernandez11},
Jim\'enez-Serra et al.~\citeyear{jimenez14}, Henshaw et al.~\citeyear{henshaw14}, Barnes et 
al.~\citeyear{barnes16}), 
both with single-dish telescopes and interferometers, allowing us to put in relation the 
\r1415\ ratios with the core properties. In particular, because each core is classified as 
starless or star-forming within each cloud, we are also able to compare the \r1415\ ratios with the
presence/absence of protostellar activity. Observations and data reduction are described in Sect.~\ref{obs}.
The results are shown in Sect.~\ref{res}, and discussed in Sect.~\ref{discu}. The main findings
of this work, and the implications for follow-up studies are presented in Sect.~\ref{conc}.

\section{Observations and data reduction}
\label{obs}

Observations have been performed with the Atacama Large Millimeter Array (ALMA) 
towards the sample of IRDCs described in Barnes et al. in prep. (hereafter paper I).
The project IDs are 2017.1.00687.S and 2018.1.00850.S (PI: A.T. Barnes). 
We refer to that work for the complete source sample description and for any observational 
detail. In this paper we analyse only the four IRDCs detected in \N15 (1--0) listed in 
Table~\ref{tab:sources}, in which we also give some source properties, as well as 
important observational parameters (angular resolution, spectral resolution, 1$\sigma$ 
rms noise in the final clean cubes). Spectroscopic parameters of the two lines analysed,
i.e. \H\ and \N15(1--0), are reported in Table~\ref{tab:spectroscopic}, together with relevant
observational parameters (i.e.~sensitivity and spectral/angular resolution). 
Calibration, imaging, and deconvolution of the interferometric data were performed with the 
Common Astronomy Software Calibration ({\sc casa}) package\footnote{CASA is developed by an 
international consortium of scientists based at the National Radio Astronomical Observatory (NRAO), 
the European Southern Observatory (ESO), the National Astronomical Observatory of Japan (NAOJ), 
the Academia Sinica Institute of Astronomy and Astrophysics (ASIAA), the CSIRO division for 
Astronomy and Space Science (CASS), and the Netherlands Institute for Radio Astronomy 
(ASTRON) under the guidance of NRAO. See https://casa.nrao.edu} (McMullin et al.~\citeyear{mcmullin07}),
using the {\sc casa-pipeline} (version: 5.4.0-70).
Continuum subtraction was performed by taking the line-free channels around the lines in each
individual spectral window (the results of the pipeline were checked), and subtracted from the 
data directly in the {\it (u,v)}-domain.  
The maximum recoverable scale was set by the size of the smallest 7m baseline of 8.9~m, which 
corresponds to $\sim 70$\asec\ at 93.2~GHz. 
Therefore, complementary total power observations were taken to recover the zero-spacing for 
the molecular lines. These single-dish observations were also reduced using the {\sc casa-pipeline} 
tool (version: 5.4.0-70). The 12m, 7m, and total power observations were combined using the 
{\sc feather} function in {\sc casa} (version 4.7.0) with the default parameter set (i.e. effective dish 
size, single-dish scaling, and low-pass filtering of the single-dish observations).
Clean cubes extracted around 
the lines of interest were then converted into fits format and further analysed with packages 
of the {\sc gildas}\footnote{https://www.iram.fr/IRAMFR/GILDAS/} software. 
In particular, the spectra of \H\ and \N15(1--0) were extracted with {\sc mapping}, and analysed
in the way described in Sect.~\ref{res} with {\sc class}.


\begin{figure*}
\centering
{\includegraphics[width=13cm,angle=0]{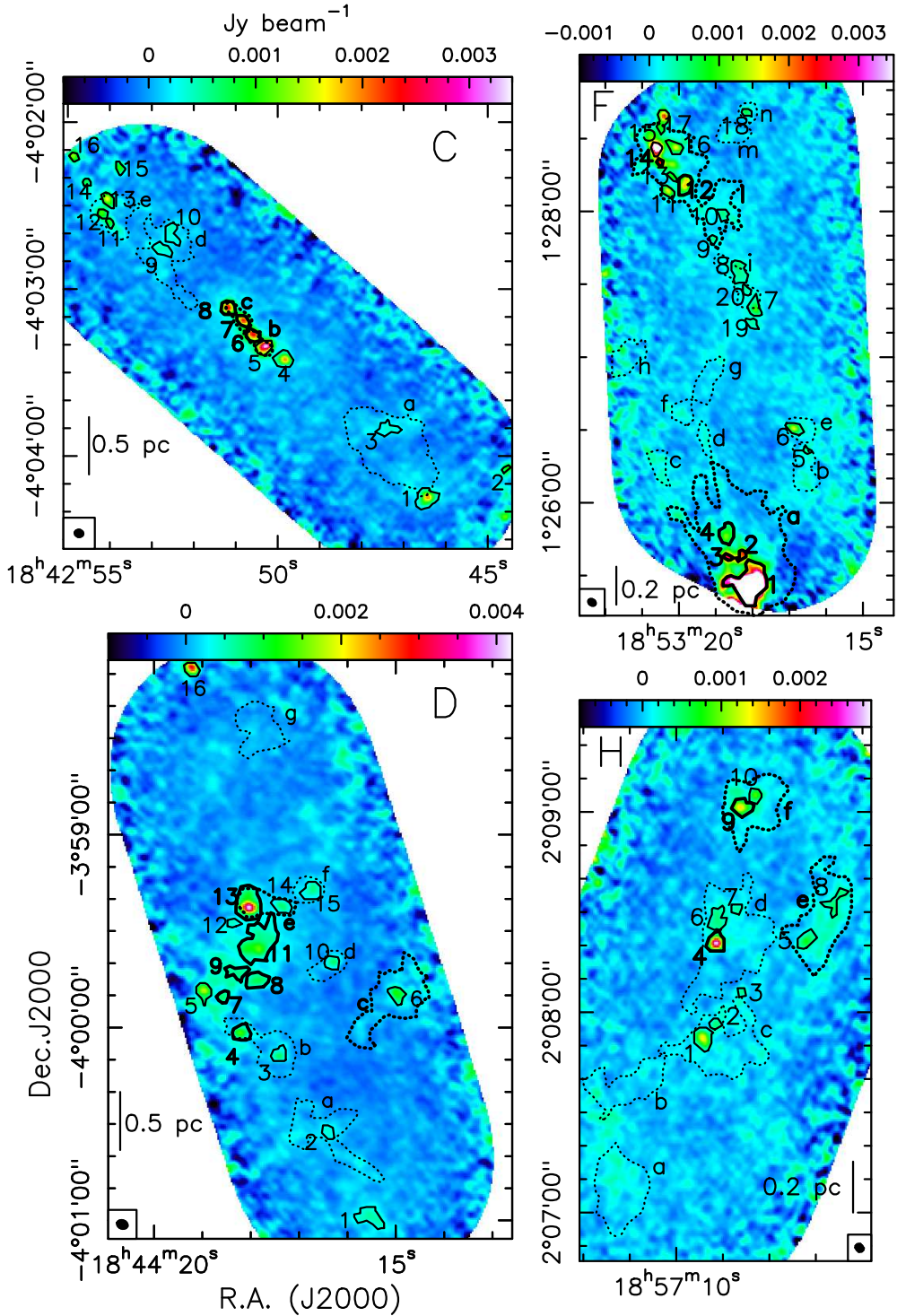}}
\caption{Continuum maps of the four target clouds, in Jy beam$^{-1}$ units. 
The solid and dotted contours illustrate the continuum and \H\ cores, respectively, derived in paper I 
from the dendrogram analysis. In both cases, the thick contours indicate the cores detected in \N15 (1--0). 
The numbers and letters correspond to the core IDs defined in the last column of 
Tables~\ref{tab:columns} and~\ref{tab:columns-letters}. 
In each frame, the ellipse in the bottom left (right for Cloud H) corner is the ALMA synthesised beam, and
the vertical solid line shows a linear scale of 0.5~pc for cloud C and D, and 0.2~pc for cloud F and H.}
\label{fig:dendro}
\end{figure*}

\begin{table*}
\begin{center}
\caption{Clouds with clear detections in \N15(1--0) and main observational parameters:
coordinates of map centre, systemic velocity ($V_{\rm sys}$), heliocentric distance ($d$),
channel width in velocity ($\Delta V$), synthesised beam ($\theta_{\rm SB}$), and root mean square noise per 
channel (1$\sigma$).}
\begin{tabular}{c c c c c c c c c c c c c c}
\hline
cloud & ID    & R.A. (J2000)   & Dec. (J2000)  & $V_{\rm sys}$ & $d$& \multicolumn{2}{c}{$\Delta V$$^{(a)}$}  & & \multicolumn{2}{c}{$\theta_{\rm SB}$} & & \multicolumn{2}{c}{1$\sigma$} \\
          &         & ${\rm h}:{\rm m}:{\prime \prime}$ & ${\circ}:{\prime}:{\prime\prime}$  & \kms\  & kpc & \multicolumn{2}{c}{\kms} & & \multicolumn{2}{c}{\asec} & & \multicolumn{2}{c}{mJy beam$^{-1}$}    \\
          \cline{7-8} \cline{10-11} \cline{13-14} \\
          &         &                                                      &                                                    &  &         & \H\ & \N15\       &               & \H\ & \N15\            &             & \H\ & \N15\             \\
\hline
C  & G028.37+00.07 & 18:42:50.03 & --04:03:23.4 & 78 & 5.0 & $\sim 0.05$ & $\sim 0.1$ & & 3.5\asec$\times 3.1$\asec & 3.5\asec$\times 3.2$\asec & & $\sim 11$ & $\sim 9$ \\
D  & G028.53--00.25 & 18:44:16.94 &  --03:59:40.8& 86 & 5.7 & $\sim 0.05$ & $\sim 0.1$ & & 3.8\asec$\times 3.0$\asec & 3.8\asec$\times 3.1$\asec & & $\sim 12$ & $\sim 9$ \\
F   & G034.43+00.24 & 18:53:18.43 & +01:27:13.5 & 57 & 3.7 & $\sim 0.05$ & $\sim 0.1$ & & 3.9\asec$\times 3.1$\asec & 4.0\asec$\times 3.1$\asec & & $\sim 11$ & $\sim 9$ \\
H  & G035.39--00.33 & 18:57:09.42 & +02:08:05.4 & 43 & 2.9 & $\sim 0.05$ & $\sim 0.1$ & & 3.4\asec$\times 3.0$\asec & 3.4\asec$\times 3.1$\asec & & $\sim 12$ & $\sim 9$\\ 
\hline
\end{tabular}
\label{tab:sources}
\end{center}
$^{(a)}$ the spectral resolution of the clean cubes corresponds to twice the channel spacing ($\Delta V$); \\
\end{table*}

\begin{table}
\begin{center}
\caption{Spectroscopic parameters of the observed lines: rest frequency ($\nu_0$), energy of the 
upper level ($E_{\rm u}$), degeneracy of the upper level ($g_{\rm u}$), Einstein coefficient for spontaneous 
emission ($A_{\rm ul}$).}
\small
\tabcolsep 0.1cm
\label{tab:spectroscopic}
\begin{tabular}{rccccc}
\hline
   transition  & hyperfine       & $\nu_0$ & $E_{\rm u}$$^{(a)}$ & $g_{\rm u}$$^{(a)}$ & $A_{\rm ul}$$^{(a)}$ \\
                    &   component       & MHz               &   K                          &                           & s$^{-1}$  \\
                    \hline
  N$_2$H$^+$ (1--0) & F$_1 = 1-1$ & 93171.88$^{(a)}$  & 4.47 & 9  & 3.63$\times 10^{-5}$  \\
                                 & F$_1 = 2-1$ & 93173.70$^{(a)}$  &   & 15 &   \\
                                 & F$_1 = 0-1$ & 93176.13$^{(a)}$ &   & 3 &   \\
 N$^{15}$NH$^{+}$ 1--0  & F $ = 1-1$ & 91204.26$^{(b)}$ & 4.33 & 3 & 3.40$\times 10^{-5}$ \\
                                      & F $ = 2-1$ & 91205.99$^{(b)}$ &  & 5 &   \\
                                     &  F $ = 0-1$ & 91208.52$^{(b)}$ &  & 1 & \\
           \hline
\end{tabular}
\end{center}
$^{(a)}$ from the Cologne Database for Molecular Spectroscopy (CDMS; Endres et al.~\citeyear{endres}); \\
$^{(b)}$ from Dore et al.~(\citeyear{dore09}); \\
\end{table}

\section{Results}
\label{res}

The \N15(1--0) has been detected towards several positions in the four clouds in Table~\ref{tab:sources}
with signal-to-noise ratio in the range $\sim 3-7$. The \H (1--0) line was clearly detected with 
excellent signal to noise ($\geq 20$) in all clouds (see Figs.~\ref{fig:spectra-cloudC} -- \ref{fig:spectra-cloudH})
after smoothing to a uniform spectral resolution of $\sim 0.4$~\kms. 
Due to the faintness of the \N15\ lines, we do not show integrated maps of the \N15\ emission. 

The faintness of the \N15\ emission, and the large angular extension and morphological complexity
of the target clouds, suggested us to focus the \r1415\ analysis first on the 3~mm 
continuum sources identified in each IRDC using the 2D dendrogram
method (Rosolowsky et al.~\citeyear{rosolowsky08}). The parameter set and method for 
determination of the continuum dendrogram structure is described in Paper I.  
The continuum sources are shown in Fig.~\ref{fig:dendro}.
We point out that the choice of the input parameters in the dendrogram 
analysis does not significantly affect the identification of the cores as demonstrated in paper I.
These sources are well characterised from the dynamical and evolutionary point of view, and hence
will allow us to discuss the \r1415\ ratio based on the core properties in Sect.~\ref{discu}. 
In particular, the cores have been classified as starless or star-forming based on the identification
of an infrared source inside the core boundaries (paper I).

We also extracted and analysed spectra from the \H\ intensity peaks to derive the \r1415\ ratio 
towards these peaks.
These have been identified here by conducting the dendrogram analysis on the (2D) integrated 
intensity maps of \H\ (see paper I). The following set of parameters are used for determination of the dendrogram 
structure: {\sc min\_value} = 30\,$\sigma$ $\sim$ 4.5 K kms (the minimum intensity considered 
in the analysis); {\sc min\_delta} = 30\,$\sigma$(the minimum spacing between isocontours); 
{\sc min\_pix} = 3\,beam $\sim$ 150 pixels (the minimum number of pixels contained within 
a structure). Given the complex and extended morphology of the \H\ emission, more stringent 
dendrogram criteria were imposed compared to the continuum. This parameter set produced 
a simplistic dendrogram hierarchy, which was preferred as this work focuses on the detection 
and analysis of the weak $^{15}$N emission, as opposed to a rigorous classification of the 
structures seen in \H\ emission.

\begin{figure}
\centering
{\includegraphics[width=8.5cm,angle=0]{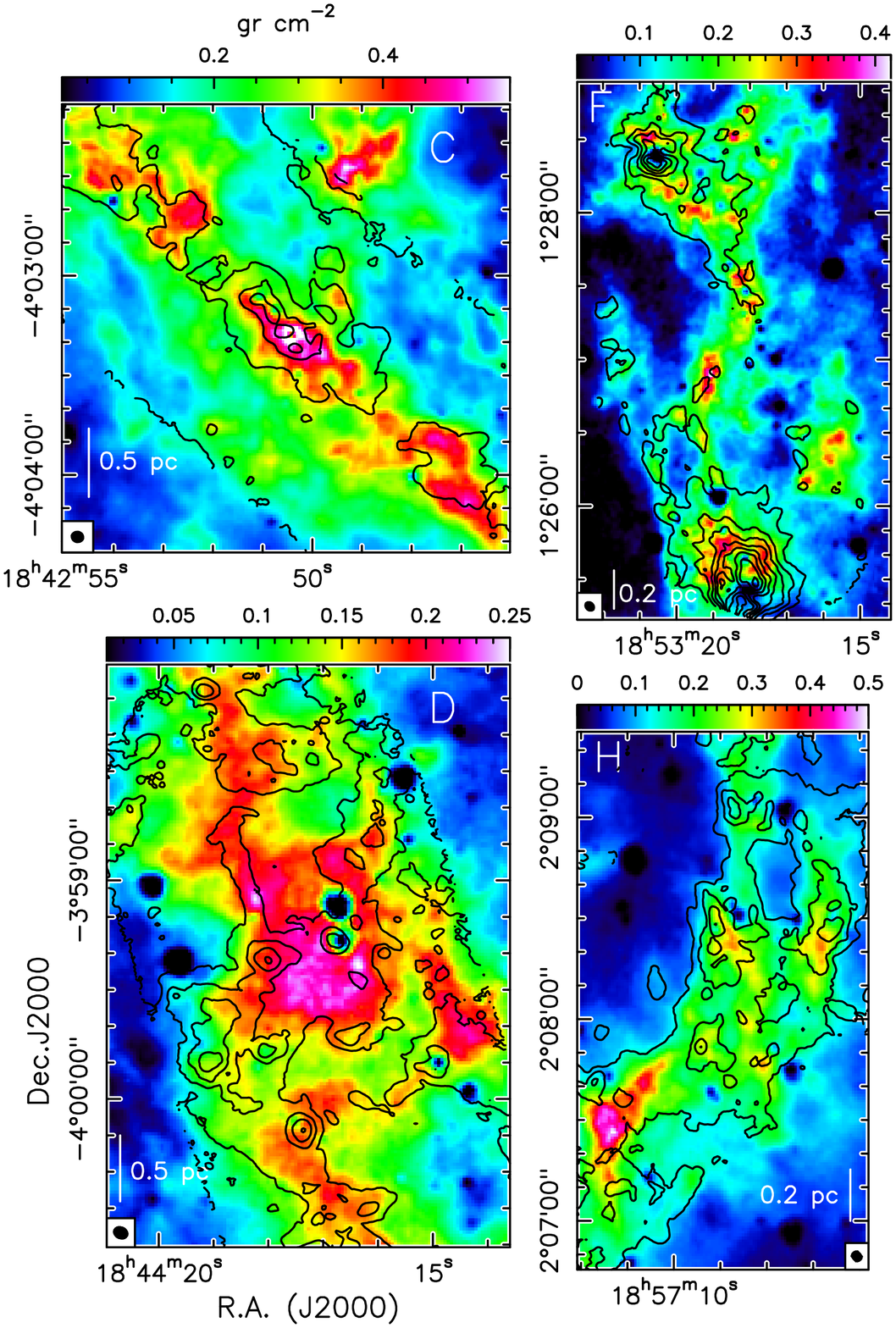}}
\caption{Mass (H$_2$) surface density maps, $\Sigma$, of the four target clouds (in gr cm$^{-2}$ units,
colour scale, Kainulainen \& Tan~\citeyear{ket13}) superimposed on the \H (1--0) line emission peak (solid
contours), in brightness temperature units, obtained with ALMA (paper I). 
In each frame, first contour and step are the $3\sigma$ rms level of the \H(1--0) peak map
(2.7~K for Cloud C; 1.4 for Cloud D; 2.7~K for Cloud F; 1.5~K for Cloud H), 
the ellipse in the bottom left (right for Cloud H) corner is the ALMA synthesised beam, and
the vertical solid line shows a reference linear scale.}
\label{fig:mir}
\end{figure}

For simplicity, in the following we will call "cores" the \H\ intensity peaks identified in this way 
as well. These \H\ cores are typically larger (i.e. more diffuse) than the continuum cores, 
due to the difference in dendrogram parameters. The \H\ emission
morphology is shown in Fig.~\ref{fig:mir}, in which we plot for each cloud the peak intensity 
of the \H (1--0) line superimposed on the mass (H$_2$) surface density map, $\Sigma$, 
obtained from combined Spitzer near- and mid-infrared extinction maps (Kainulainen \& Tan~\citeyear{ket13}): 
overall, the $\Sigma$ morphology matches well the \H\ emission peak in Cloud C, F, 
and H, even though the peaks of the two tracers coincide relatively well only in Cloud C,
while in cloud H and F some \H\ peaks are not coincident with the $\Sigma$ peaks.
This apparent disagreement between the \H(1--0) intensity peak and $\Sigma$ peaks,
especially in Cloud D and partly also in Cloud H and F, could be due to several causes affecting 
the \H\ line profile in the regions with high extinction. 
Indeed, as we will discuss in Sect.~\ref{n2hp-cont}, the \H\ spectra show a very complex 
profile suggestive of multiple velocity components, self-absorption, and high-optical depth effects,
in particular towards Cloud D and H.

The contours used for the extraction of the spectra are shown in Fig.~\ref{fig:dendro}, where we
can note that several continuum cores are superimposed, or totally included, in the \H\ cores. 
However, the \H\ cores are usually more extended, and in some cases they are not detected in 
the 3~mm continuum, and vice-versa. A similar feature was found by Di Francesco et al.~(\citeyear{difrancesco04}) in 
Oph A, and interpreted as being due to the fact that \H\ (1--0) is tracing the cold gas better than the dust 
continuum emission, which is biased toward warm regions. Therefore, our "double" analysis allows 
us to investigate if the \r1415\ ratios change from the compact continuum cores to the more diffuse 
\H\ ones.

We will present first the results of the data analysis of the spectra extracted from the continuum cores 
in Sect.~\ref{cont}, and then we will show those obtained from the \H\ cores in Sect.~\ref{n2hp}, also
comparing the two results.

\subsection{Spectra towards the 3~mm continuum cores}
\label{cont}

\subsubsection{Extraction of the spectra and line detection}
\label{detection-cont}

We extracted \N15(1--0) spectra from the continuum sources identified in paper I, and detected significant
emission towards three cores in Cloud C ($\sim 19\%$), six cores in Cloud D ($\sim 38\%$), 
six cores in Cloud F ($\sim 30\%$), and two cores in Cloud H ($\sim 20\%$). 
The contours used for the extraction of the spectra are shown in Fig.~\ref{fig:dendro}, where
we highlight the cores detected in \N15. The integrated flux density, $F_{\nu}$, of each 
spectrum has been converted to brightness temperature units via the formula:
\begin{equation}
T_{\rm B}(K) \simeq 1222 F_{\nu}{\rm (mJy)}/[ \nu^2{\rm (GHz)}\Theta^2 (\prime \prime)] \;,
\label{eq:conversion}
\end{equation}
where $\nu$ is the rest frequency of the transition (see Table~\ref{tab:spectroscopic}) and $\Theta$ 
is the observed angular equivalent diameter (i.e. the diameter of the equivalent circle) of 
each core (paper I). Therefore, $T_{\rm B}$ is an average brightness temperature over the 
angular surface of each core. We stress that Eq.(\ref{eq:conversion}) is valid also for non-spherical
sources as long as $\Theta$ is the source equivalent angular diameter. Following the same approach, 
we will have extracted and analysed spectra from the \H\ cores, which will be discussed in Sect.~\ref{n2hp}

The spectra in $T_{\rm B}$ units are shown in Figs.~\ref{fig:spectra-cloudC} -- \ref{fig:spectra-cloudH}. The 
\N15\ emission is preferentially detected towards cores with strong continuum emission (cores C2c3/5, C2c4, 
C2c6 in Cloud C; core D7c3 in Cloud D; Cores F1c1, F1c2, F1c3, F4c5, F4c7 in Cloud F; cores H3c3 and H5c3 
in Cloud H), and are usually located in the central part of the cloud, with the exception of the cores in Cloud F, 
located at the southern and northern edge of the cloud (Fig.~\ref{fig:dendro}). 

\subsubsection{Line fits: \N15 (1--0)}
\label{n15nhp-cont}

The \N15 (1--0) line possesses hyperfine structure due to the nuclear spin of the $^{14}$N 
nucleus, which splits the transition in three components well separated in frequency, with the strongest 
one at the centre of the spectrum, and two fainter satellites. The frequencies of the components and
relevant spectroscopic parameters are given in Table~\ref{tab:spectroscopic}. 
In all but four cores, D5c5/6/7, F1C3, F4c5, and F4c7, the hyperfine structure cannot be fitted because 
the satellites are undetected. Therefore, the main components have been fitted with 
a single Gaussian, except in the four cores mentioned above, for which we have fitted simultaneously all components
using the CLASS software (method hfs in CLASS): the method considers that the velocity separation of 
the components is fixed to the laboratory values, and that all components have the same excitation temperature 
(\Tex) and the same line full width at half maximum (FWHM).
 
By fitting a single Gaussian to the lines in which only the main hyperfine component is detected, 
we are underestimating the total line integrated area. To compute the total column density we
have appropriately derived the total integrated intensity from our partial estimates as explained in 
Sect.~\ref{coldens-cont}.
 
\subsubsection{Line fits: \H (1--0)}
\label{n2hp-cont}

The \H\ (1--0) has also hyperfine structure with 15 components, of which seven are 
resolved in velocity in quiescent cores (see e.g.~Caselli
et al.~\citeyear{caselli95}). However, in many observed spectra, including ours, the  
line widths are much larger than the separation in velocity of the components, which are thus
grouped in three spectral features (Figs.~\ref{fig:spectra-cloudC} -- \ref{fig:spectra-cloudH}).
Moreover, the profile is often complex and difficult to fit even considering the hyperfine 
structure. Such complexity can be attributed to multiple velocity components, known 
to be present in the targets (e.g.~Henshaw et al.~\citeyear{henshaw13},~\citeyear{henshaw14}; 
Barnes et al.~\citeyear{barnes18}), and/or to optical depth effects and self-absorption.

Let us discuss now in detail the profiles of the \H\ (1--0) lines in each cloud. We examine 
in particular the spectra of the cores also detected in \N15 (1--0), from which we will compute
the \r1415\ ratios:
\begin{itemize}
\item [-]Cloud C: all spectra show a main component centred at (or very close to) the systemic 
velocity of the cloud ($\sim 78$~\kms), and a fainter red-shifted emission, most apparent in, e.g.,
C1c1, C2c1 and C2c2 (Fig.~\ref{fig:spectra-cloudC}). The red-shifted emission is also clearly apparent 
in the three cores detected in \N15, i.e. C2c3/5, C2c4, and C2c6. In particular, in C2c3/5 and C2c4 the 
central velocity of the main component is displaced by $\sim 1$~\kms\ with respect to the \N15 (1--0) line peak, 
suggesting that self-absorption could cause the asymmetric profile of \H(1--0). Moreover, the FWHM
of the \N15 (1--0) lines are narrower that those of \H (1--0), which again could be due either to
self-absorption in the main isotopologue line, or to multiple overlapping velocity features, undetected 
in \N15 (1--0) because too faint. To understand which of these two phenomena produces more
likely the observed line features in \H\ and \N15, we have first
performed a fit with two velocity components in \H, which gives good results in the three cores.
Then, we have compared the \N15\ line velocity peak with that of the two candidate
velocity features in the isolated hyperfine component F$_1 = 0-1$ of \H(1--0): we found that the 
peak of the strongest velocity feature in \H\ coincides within the uncertainties with the \N15\ velocity 
peak, favouring, for \H (1--0), the multiple velocity component hypothesis with respect to self-absorption;
\item [-] Cloud D: this is the cloud with the most complex spectral features. Fits to the line profiles
even including multiple velocity components and high optical depths provide high residuals in the 
six cores detected in \N15, i.e. D6c4, D6c6, D7c1, D7c2, D5c5/6, and D8c1. 
As for Cloud C, the displacement of $\sim 1$~\kms\ or more between the peaks 
of the \N15 (1--0) and the velocity peak of the main component of \H (1--0) suggests 
that self-absorption can be important. But multiple velocity components, undetected
in \N15 (1--0), cloud also contribute to the total line profile. As for Cloud C, we have compared
the peak velocity of \N15(1--0) with the possible velocity features in the F$_1 = 0-1$
component of \H(1--0), and found that in three cases, i.e. D6c6, D7c1, and D7c2, 
the peak velocity of the \N15(1--0) sits in between the peaks of the two velocity feature 
candidates seen in \H\ (see Fig.~\ref{fig:spectra-cloudD}).
Therefore, in these cores the \H\ asymmetric line profile is likely due to self-absorption.
In the other three cores, i.e.~D6c4, D5c5/6, and D8c1, the profile may be due to two
velocity features;
\item [-] Cloud F: The lines can be fitted with a single velocity component in all spectra except
that in F1c5, F1c6, F2c1, and F2c2, where there seem to be at least two velocity components
(Fig.~\ref{fig:spectra-cloudF}). Among the five cores detected in \N15 (1--0), i.e. F1c1, 
F1c2, F1c3, F1c4, F4c5 and F4c7, only F1c2 and F1c3 show hints of a secondary blue-shifted
velocity feature. However, in both cases the \N15\ line peaks sits neither in between the
possible two velocity features nor on one of them precisely, and hence one cannot establish which
is the (main) cause of the asymmetric profile in \H. Nevertheless, these possible secondary velocity
features contribute to the integrated intensity by less than $20\%$ (\ref{fig:spectra-cloudF}),
and hence does not affect strongly the integrated intensity estimate;
\item [-] Cloud H: the line profiles are very different among the various cores (\ref{fig:spectra-cloudH}), 
going from single velocity components with moderate (H3c3) or high (H5c3, H5c1) optical depth, to multiple 
velocity features (H2c1, H4c1, H3c1). Again, the profile of the isolated F$_1 = 0-1$ hyperfine component is the 
least affected, and is well fitted by a single Gaussian in the two cores detected in \N15, i.e.~H3c3 and H5c3.
\end{itemize}


\subsubsection{Derivation of the total column densities}
\label{coldens-cont}

The complexity of the \H (1--0) line profiles, clearly seen in the spectra of Figs.~\ref{fig:spectra-cloudC} 
-- \ref{fig:spectra-cloudH}, and described in Sect.~\ref{n2hp-cont}, makes it difficult to obtain fits with low
residuals. Therefore, we have estimated the \H\ total column density,
$N_{\rm tot}$(\H), from the integrated intensity of the hyperfine component F$_1 = 0-1$, which is well
separated in velocity from the others and intrinsically the faintest one, and hence potentially the least 
affected by optical depth effects, self-absorption, or contribution of secondary faint velocity features. 
This was then converted to the total integrated intensity of the line by multiplying it by 9, which is 
the relative strength of the sum of all components with respect to the F$_1 = 0-1$ one 
(see Table~\ref{tab:spectroscopic}).
Then, the total column density was calculated from the total integrated intensity using Eq.~(A4) of 
Caselli et al.~(\citeyear{caselli02}), which assumes optically thin emission and same (and constant) 
excitation temperature, \Tex, for all rotational transitions (and all hyperfine components within
each transition). 

Obviously, this simplified approach provides column density estimates that need to be
taken with caution, because the spectra of \H\ in some cases show hints of multiple
velocity features and/or self-absorption also in the F$_1 = 0-1$ component 
(see Figs.~\ref{fig:spectra-cloudC} -- \ref{fig:spectra-cloudH}), as discussed in 
Sect.~\ref{n2hp-cont}. In particular, when we could determine that the profile is due 
to multiple velocity features, we have integrated only the strongest velocity component,
(e.g. the three cores in Cloud C, or D6c4, D5c5/6 and D8c1 in cloudD, 
see Sect.~\ref{n2hp-cont}); when the line profile was attributed more 
likely to self-absorption (e.g. in the cores D6c6, D7c1 and D7c2), we have integrated the 
whole component, bearing in mind that this will provide a lower limit on $N_{\rm tot}$(\H),
and hence a lower limit on the associated \r1415\ as well. In cores undetected in \N15, one cannot 
establish if the line profile of \H\ is most likely due to multiple velocity components or to self-absorption,
except in the cases in which velocity features well-separated in velocity are found (e.g.~in
H2c1 and H3c1). In the latter case, we have performed a fit assuming multiple velocity features,
identified the strongest one (even in case of large residuals), and taken the integrated 
intensity of this velocity feature only.

Because the adopted approach does not allow to estimate the excitation temperature, \Tex, 
we have assumed a range of \Tex\ which can be considered typical or reasonable for IRDC
cores, i.e.~$\sim 10 -50$~K. This temperature range includes the dust temperature estimates
obtained from Herschel for all clouds (see paper I), and the kinetic temperature estimates
obtained in Cloud H ($\sim 10-20$~K, Sokolov et al.~\citeyear{sokolov17}).
The corresponding range of values for $N_{\rm tot}$(\H) is $\sim 0.8 - 14 \times 10^{13}$ \cmq\ 
at \Tex = 10 K, and $\sim 2.4 - 41 \times 10^{14}$ \cmq\ at \Tex = 50 K.
The results are shown in Table~\ref{tab:columns}.

For \N15, we have followed the same method as that used for the \H\ lines:
for the spectra in which the main hyperfine component only is detected, i.e. the F$=2-1$ one 
(see Table~\ref{tab:spectroscopic}), we have estimated the total integrated intensity by dividing 
the integral of this component for its relative strength (i.e.~$\sim 0.555$, Table~\ref{tab:spectroscopic}).
The approach is appropriate for optically thin lines. We note that we cannot measure directly the 
optical depth of the lines because we detected the hyperfine structure only towards two cores, 
and even in these cases the measured
$\tau$ is very uncertain (error on $\tau$ larger than $\tau$). However, the assumption of optically thin 
lines is justified by the faint emission and the non-detection of the satellites, and supported 
by direct measurements in protostellar cores in which $\tau$ of \N15(1--0) is comparable to or 
smaller than $\sim 0.1$ (e.g.~towards OMC--2 FIR4, Fontani et al.~\citeyear{fontani20}). 
As excitation temperature, we have used the same one assumed for \H. 

We find $N_{\rm tot}$(\N15) in the range $\sim 1.3 - 7 \times 10^{11}$ \cmq\ at \Tex = 10 K, 
and $\sim 4-19 \times 10^{11}$ \cmq\ at \Tex = 50 K.
We point out that even though the assumed temperature affects the column densities of \H\ and
\N15, it does not affect at all the \r1415\ ratio, as we will see in Sect.~\ref{1415-cont}.

Finally, in the cores undetected in \N15, we have evaluated the upper limit on $N_{\rm tot}$(\N15) 
from the upper limit on the integrated intensity, estimated through:
\begin{equation}
\int T_{\rm B}d{\rm v} = \frac{1}{0.555}3\sigma\frac{\sqrt{\pi}}{2\sqrt{{\rm ln2}}}{\rm FWHM}\;,
\label{eq:upper}
\end{equation}
which expresses the integral in velocity of a Gaussian line with peak intensity given by the $3\sigma$ rms 
in the spectrum. The factor 0.555, which is the relative strength of the F$=2-1$ hyperfine component, is 
introduced to convert the upper limit of this component to the upper limit of the total one.
The assumed FWHM is the mean value measured from the \N15\ lines detected in each cloud.

The \H\ and \N15\ column densities have been calculated assuming that \H\ and \N15 (1--0) have the 
same \Tex\ based on the assumption that the two transitions have very similar critical densities, and 
hence similar excitation conditions. However, let us discuss this assumption better. Based on a non-LTE analysis, 
Hily-Blant et al.~(\citeyear{hily-blant13a}) found differences in \Tex\ for lines with the same quantum 
numbers of the different isotopologues of HCN. But these differences are in all (but one) cases below 
$\sim 10\%$, indicating that a significantly different \Tex\ for lines with the same quantum numbers is 
unlikely for isotopologues of the same species. 
Regarding the possibility that different hyperfine components of the same isotopologue can have a different 
\Tex, Daniel et al.~(\citeyear{daniel06}) showed that high optical depths in \H\ (1--0) could indeed cause 
deviations from the line profile expected when each component has the same excitation temperature. 
However, both theoretical (Daniel et al.~\citeyear{daniel06}) and observational (Caselli et al.~\citeyear{caselli95}) 
works show that \Tex\ of the component analysed in our work, i.e. the F$_1 = 0-1$ one,
would deviate from the local thermodynamic equilibrium value by 10-15$\%$ at most even in the
high optical depth case, and only at H$_2$ volume densities below $10^{5}$~\cmc\ (see Fig.~6 in 
Daniel et al.~\citeyear{daniel06}). In our cores, the average H$_2$ volume density is 
$n_\mathrm{H_2} = 7.6_{-2.6}^{+5.2}\times10^5$\,cm$^3$, or when using the background subtracted 
mass $n^\mathrm{b}_\mathrm{H_2} = 3.4_{-1.8}^{+2.4}\times10^5$\,cm$^3$
(paper I), for which the predicted deviations from the equilibrium \Tex\ is negligible (Fig.~6 in Daniel et 
al.~\citeyear{daniel06}). Therefore we are confident that hyperfine anomalies are not affecting 
significantly the \Tex\ of the analysed component. 

\subsection{Spectra towards the \H\ cores}
\label{n2hp}

For completeness, we have extracted and analysed the spectra associated with the cores 
identified through the \H\ emission. These were obtained with the dendrogram analysis like
the 3~mm continuum cores in paper I, and are shown in Fig.~\ref{fig:dendro}, labelled following
a method similar to the one used in paper I for the continuum cores: we identify the closer core in
Butler \& Tan~(\citeyear{bet12}), to which we add the label "n2hp", followed by
 a sequential number in case more than one core can be associated with the same 
 Butler \& Tan~(\citeyear{bet12}) core. Most of these cores are more 
extended than the continuum ones, which are included in the \H\ contours especially in clouds
C, D and H. Therefore, this analysis allows us to investigate if the \N15\ emission is more 
extended, or, to say it in another way, "detectable" also away from the continuum cores.
The method adopted to extract the spectra from these cores, convert them from flux density
to $T_{\rm B}$ unit, fit them, and derive the column densities, are identical to those described
in Sect.~\ref{cont}. 

In Fig.~\ref{fig:dendro} we highlight the cores in which \N15\ has been detected with signal-to-noise 
ratio $\geq 3$. In Cloud C, the two detected cores, C2-n2hp-1 and C2-n2hp-2, include the continuum 
cores already detected in \N15, i.e. C2c3/5, C2c4, and C2c6; therefore, we have not found "new" detections.
The same result is found in Cloud F, in which again \N15(1--0) is detected only towards the two cores
(F1-n2hp-1 and F4-n2hp-1) that include in their contours all the continuum cores already detected
in this line. 
In Cloud D, we have detected \N15\ towards D5-n2hp-1 and D7-n2hp-1: D7-n2hp-1 includes the already 
detected continuum core D8c1, while D5-n2hp-1 contains the undetected continuum core D5c7, 
but D5-n2hp-1 is much more extended than D5c7. Hence, in this case the \N15\ emission arises
from the diffuse envelope of D5c7. Apart from this case, overall in clouds C, D, and F the detections 
in the continuum and \H\ cores are always consistent. 

\begin{figure}
\centering
\includegraphics[width=7cm,angle=0]{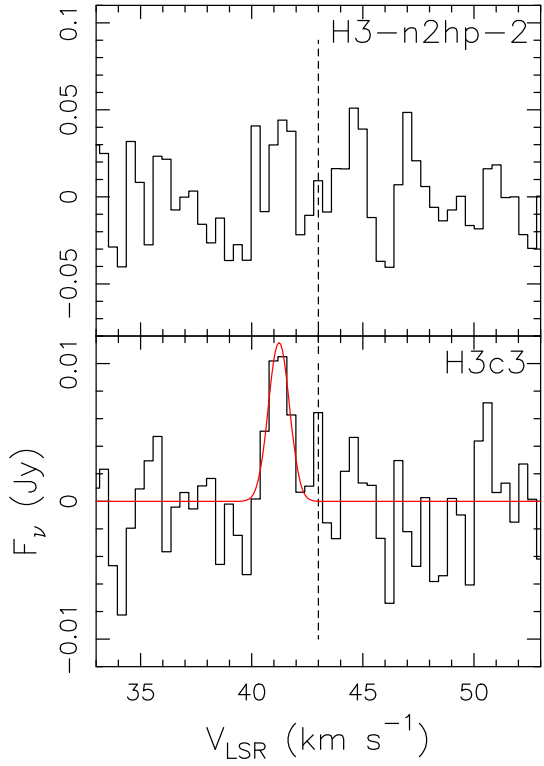}
\caption{Comparison between the spectra of \N15(1--0) extracted in flux density units of cores 
H--d (top) and H3c3 (bottom). H3c3 is included in H3-n2hp-2, but the higher rms noise in H--d prevents the 
detection of \N15\ (1--0) at about 41.2~\kms\ towards H3c3. The dashed vertical line indicates 
the systemic velocity of Cloud H of 43~\kms.}
\label{fig:H-d}
\end{figure}

The case of Cloud H is more complicated. In this cloud we detect \N15\ towards two cores: H2-n2hp-1, 
which includes the much more compact undetected continuum cores H2c1 and H3c1, and H5-n2hp-1, which 
includes the continuum core H5c3, detected in \N15. Hence, the detection in H2-n2hp-1 is due to emission in the
extended envelopes of H2c1 and H3c1, as in D5-n2hp-1. There is also a peculiar case: H3-n2hp-2, which is undetected 
in \N15\ but includes the compact continuum core H3c3 detected in \N15. Therefore, the non-detection in H3-n2hp-2 appears
inconsistent at first glance with the detection towards H3c3. In this latter, \N15 (1--0) is displaced by $\sim 1.8$~\kms\ from 
the systemic velocity of the cloud (see Fig.~\ref{fig:H-d}). This detection could hence be doubtful. 
However, a displacement of about 1~\kms\ is seen also in the \H\ (1--0) line (see Fig.~\ref{fig:spectra-cloudH}).
Furthermore, and most importantly, the noise in the H3-n2hp-2 spectrum is about 5 times higher than that in 
the H3c3 one (see Fig.~\ref{fig:H-d}), and the hint of a possible line under the noise level is apparent in 
H3-n2hp-2 as well. Therefore, we conclude that the detection of \N15\ towards H3c3, detected with
signal-to-noise of 5, is real.

In summary, \N15 (1--0) has been detected towards \H\ cores not previously detected in continuum 
cores only in two sources, D5-n2hp-1 and H2-n2hp-1. This result suggests that, overall, the \N15\ emission 
seems to be fainter in the more diffuse \H\ cores than in the compact continuum cores. We will discuss 
further this finding in Sects.~\ref{1415-cont} and \ref{1415-n2hp}.

The \H\ and \N15\ column densities have been computed adopting the same approach described in 
Sect.~\ref{coldens-cont}. Among the cores that show complex line profiles in \H(1--0), D7-n2hp-1 is the only
one in which the profile can be attributed to self-absorption based on the comparison between the velocity 
peak of \N15\ and the F$_1 = 0-1$ hyperfine component of \H. 
The results are shown in Table~\ref{tab:columns-letters}.
For \H, we find $N_{\rm tot}$ in the range $0.55-5.5\times 10^{13}$~\cmq\ 
when \Tex\ is $\sim 10$~K, and $1.6-16.1\times 10^{13}$~\cmq\ when \Tex\ is $\sim 50$~K. For \N15, 
we find $N_{\rm tot}$ in the range $0.8-3.3 \times 10^{11}$~\cmq\ when \Tex\ is $\sim 10$~K,
and $2.3-9.6 \times 10^{11}$~\cmq\ when \Tex\ is $\sim 50$~K.
In Table~\ref{tab:columns-letters} we also list the \D\ cores identified in Kong et al.~(\citeyear{kong17}) 
which can be associated with the \H\ cores. A few of them are indeed associated, but unfortunately 
only 4 of those are also detected in \N15, and thus the statistics is too low to discuss possible relations 
between nitrogen and deuterium fractionation.

\onecolumn
\begin{longtable}{cccccccccc}
\caption{\label{tab:columns} Total integrated line intensities ($\int T_{\rm B}d{\rm v}$) and total column densities ($N_{\rm tot}$)
of \H\ and \N15\ derived from the (1--0) transition as explained in Sect.~\ref{res}, and their ratio \r1415. 
For \H(1--0) the listed $\int T_{\rm B}d{\rm v}$ is obtained dividing the integrated intensity of the isolated hyperfine component 
(i.e. F$_1$=0--1, Col.~2) by 0.1111 (relative intensity of this component with respect to the total one).
These integrated intensities are shown in Figs.~\ref{fig:spectra-cloudC}, \ref{fig:spectra-cloudD}, \ref{fig:spectra-cloudF}, 
and \ref{fig:spectra-cloudH}. 
For lines with multiple velocity features also in the F$_1$=0--1 component and detected in \N15, we used 
the integrated intensity only of the feature corresponding in velocity to that of the \N15\ line.
For \N15 (1--0), we have computed $\int T_{\rm B}d{\rm v}$ from the integrated intensity of the main hyperfine 
component divided by 0.5555 (relative intensity of this component with respect to the total one) 
for all sources except for D5c5/6, F4c5,  F4c7, for which all hyperfine components could be detected and
integrated (Col.~5). The error on the line integral
was computed according to the propagation of errors, and it is given by the expression 
$\sigma \times \Delta V \times \sqrt{n}$, where $\sigma$ is the rms in the spectrum, $\Delta V$
the velocity resolution, and $n$ the number of integrated channels). 
The errors on the column densities include the error on the line integrated intensity and the
calibration uncertainty on the absolute flux density scale of 10$\%$.
The quoted \r1415\ are derived from the column densities estimated assuming \Tex = 10~K, 
but those computed when using \Tex = 50~K are identical.
The core names in the first column are taken from paper I, while in the last column we list the 
core ID number used in Fig.~\ref{fig:dendro}.} \\
\hline
\onecolumn
source &  \multicolumn{3}{c}{\H}   & &     \multicolumn{3}{c}{\N15}  &  & \\
\cline{2-4} \cline{6-8} \\
            &  $\int T_{\rm B}d{\rm v}$ & \multicolumn{2}{c}{$N_{\rm tot}$} & & $\int T_{\rm B}d{\rm v}$ & \multicolumn{2}{c}{$N_{\rm tot}$} & \r1415  & ID \\
            &    K \kms\    &   $\times 10^{13}$\cmq\     &   $\times 10^{13}$\cmq\  &  & K \kms\       &   $\times 10^{11}$\cmq\     &   $\times 10^{11}$\cmq\    &        \\
            &                    &     $T_{\rm ex}$ = 10~K             &      $T_{\rm ex}$ = 50~K          &       &       & $T_{\rm ex}$ = 10~K & $T_{\rm ex}$ = 50~K  & \\
\hline
\endfirsthead
\caption{continued.} \\
\hline
source &  \multicolumn{3}{c}{\H}   & &     \multicolumn{3}{c}{\N15}  &  \\
\cline{2-4} \cline{6-8} \\
            &  $\int T_{\rm B}d{\rm v}$ & \multicolumn{2}{c}{$N_{\rm tot}$} & & $\int T_{\rm B}d{\rm v}$ & \multicolumn{2}{c}{$N_{\rm tot}$} & \r1415 & ID \\
            &    K \kms\    &   $\times 10^{13}$\cmq\     &   $\times 10^{13}$\cmq\  &  & K \kms\       &   $\times 10^{11}$\cmq\     &   $\times 10^{11}$\cmq\    &        \\
            &                    &     $T_{\rm ex}$ = 10~K             &      $T_{\rm ex}$ = 50~K          &       &       & $T_{\rm ex}$ = 10~K & $T_{\rm ex}$ = 50~K  & \\
\hline
\endhead
\hline
   C1c1   &  13.8(0.2)  &   1.9(0.2)   &  5.4(0.6)     &  &     $\leq 0.08$  &   $\leq 1.2$  &   $\leq 3.4$   & $\geq 160$  &     1  \\
   C1c2   &  11.2(0.1)  &   1.5(0.2)   &  4.4(0.5)     &  &     $\leq 0.30$  &   $\leq 4.2$  &   $\leq 12.4$  & $\geq 40$  &      2  \\
   C1c3   &  21.4(0.3)  &   2.9(0.3)   &  8.4(1.0)     &  &     $\leq 0.08$  &   $\leq 1.1$  &   $\leq 3.1$   & $\geq 270$ &     3  \\
   C2c1   &  25.2(0.2)  &   3.4(0.4)   &  9.9(1.1)     &  &     $\leq 0.07$  &   $\leq 1.0$  &   $\leq 2.9$   & $\geq 345$ &     4  \\
   C2c2   &  42.6(0.2)  &   5.7(0.6)   &  16.7(1.8)    &  &     $\leq 0.08$  &   $\leq 1.2$  &   $\leq 3.4$   & $\geq 490$ &      5  \\
 C2c3/5   &  37.6(0.3)  &   5.1(0.6)   &  14.8(1.6)    &  &     0.15(0.03)   &   2.1(0.7)    &   6(2)        & $240 \pm 100$  &     6  \\
   C2c4   &  46.7(0.3)  &   6.3(0.7)   &  18(2)        &  &     0.17(0.05)   &   2.4(0.9)    &   7(3)         & $260 \pm 120$ &      7  \\
   C2c6   &  34.8(0.2)  &   4.7(0.5)   &  13.7(1.5)    &  &     0.19(0.04)   &   2.6(0.9)    &   8(3)         & $180 \pm 80$ &    8  \\
 C5c3/4   &  17.2(0.2)  &   2.3(0.3)   &  6.8(0.7)     &  &     $\leq 0.09$  &   $\leq 1.3$  &   $\leq 3.7$   & $\geq 190$ &      9 \\
 C5c1/2   &  21.1(0.3)  &   2.8(0.3)   &  8.3(0.9)     &  &     $\leq 0.07$  &   $\leq 0.9$  &   $\leq 2.8$   & $\geq 300$ &     10 \\
   C6c5   &  16.9(0.3)  &   2.3(0.3)   &  6.7(0.8)     &  &     $\leq 0.13$  &   $\leq 1.8$  &   $\leq 5.2$   & $\geq 130$ &     11 \\
 C6c6/8   &  20.7(0.4)  &   2.8(0.3)   &  8.1(1.0)     &  &     $\leq 0.14$  &   $\leq 1.9$  &   $\leq 5.6$   & $\geq 150$ &    12  \\
 C6c1/2   &  19.0(0.3)  &   2.6(0.3)   &  7.5(0.9)     &  &     $\leq 0.13$  &   $\leq 1.8$  &   $\leq 5.2$   & $\geq 145$ &     13  \\
   C6c7   &  19.0(0.5)  &   2.6(0.3)   &  7.5(0.9)     &  &     $\leq 0.19$  &   $\leq 2.7$  &   $\leq 7.9$   & $\geq 100$ &     14  \\
   C6c9   &   9.5(0.5)  &   1.3(0.2)   &  3.8(0.6)     &  &     $\leq 0.16$  &   $\leq 2.3$  &   $\leq 6.7$   & $\geq 60$ &     15  \\
  C6c10   &  15.7(0.7)  &   2.1(0.3)   &  6.2(0.9)     &  &     $\leq 0.30$  &   $\leq 4.2$  &   $\leq 12.5$  & $\geq 50$ &     16 \\
  \hline
 D3c3/4   &   8.1(0.3)  &   1.1(0.14)  &   3.2(0.4)    &  &     $\leq 0.09$  &   $\leq 1.2$  &   $\leq 3.5$   & $\geq 90$ &      1 \\
   D3c5   &   6.8(0.2)  &   0.9(0.14)  &   2.7(0.3)    &  &     $\leq 0.08$  &   $\leq 1.15$ &   $\leq 3.4$   & $\geq 80$ &      2 \\
D6c1/2/3   &  20.3(0.2)  &   2.7(0.3)   &   8.0(0.9)    &  &     $\leq 0.07$  &   $\leq 1.0$  &   $\leq 3.1$   & $\geq 260$ &      3 \\
   D6c4   &  18.8(0.3)  &   2.5(0.3)   &   7.4(0.9)    &  &     0.15(0.05)   &   2.0(0.9)    &   6(3)         & $120 \pm 70$ &      4 \\
   D6c5   &  29.2(0.4)  &   3.9(0.4)   &  11.5(1.3)    &  &     $\leq 0.13$  &   $\leq 1.9$  &   $\leq 5.5$   & $\geq 210$ &      5  \\
   D5c7   &  26.4(0.4)  &   3.6(0.4)   &  10.4(1.2)    &  &     $\leq 0.11$  &   $\leq 1.6$  &   $\leq 4.7$   & $\geq 220$ &      6 \\
   D6c6   &  24.0(0.5)  &   3.2(0.4)   &  9.5(1.1)     &  &     0.10(0.05)   &   1.4(0.8)    &   4(2)         & $230 \pm 160$ &      7 \\
   D7c1   &  21.2(0.2)  &   2.9(0.3)   &  8.4(0.9)     &  &     0.12(0.04)   &   1.7(0.7)    &   5(2)         & $170 \pm 90$ &     8 \\
   D7c2   &  21.6(0.2)  &   2.9(0.3)   &  8.5(0.9)     &  &     0.16(0.05)   &   2.2(0.9)    &   6(3)         & $130 \pm 70$ &      9 \\
   D5c7   &  14.2(0.2)  &   1.9(0.2)   &  5.6(0.6)     &  &     $\leq 0.07$  &   $\leq 0.9$  &   $\leq 2.7$   & $\geq 210$ &    10 \\
 D5c5/6   &  16.6(0.1)  &   2.2(0.2)   &  6.5(0.7)     &  &     0.09(0.04)   &   1.3(0.7)    &   4(2)         & $170 \pm 100$ &     11 \\
   D7c3   &  17.2(0.3)  &   2.3(0.3)   &  6.8(0.8)     &  &     $\leq 0.07$  &   $\leq 1.0$  &   $\leq 3.0$   & $\geq 230$ &     12 \\
   D8c1   &  28.9(0.1)  &   3.9(0.4)   &  11.4(1.2)    &  &     0.10(0.02)   &   1.4(0.4)    &   4.0(1.3)     & $280 \pm 120$ &     13 \\
   D5c3   &  27.0(0.2)  &   3.6(0.4)   &  10.6(1.2)    &  &     $\leq 0.06$  &   $\leq 0.9$  &   $\leq 2.6$   & $\geq 400$ &     14 \\
   D5c1   &  19.3(0.2)  &   2.6(0.3)   &  7.6(0.8)     &  &     $\leq 0.06$  &   $\leq 0.9$  &   $\leq 2.6$   & $\geq 290$ &     15 \\
 D9c1/2   &  21.8(0.6)  &   2.9(0.4)   &  8.6(1.1)     &  &     $\leq 0.17$  &   $\leq 2.4$  &   $\leq 7.1$   & $\geq 120$ &     16 \\
 \hline
   F1c1   &  104.2(0.3) &   14.0(1.4)  &   41(4)       &  &     0.26(0.03)   &   3.6(0.8)    &   11(2)        & $390 \pm 130$ &      1 \\
   F1c2   &  89.6(0.5)  &   12.1(1.3)  &   35(4)       &  &     0.21(0.06)   &   3.0(1.1)    &   9(3)         & $400 \pm 190$ &     2 \\
   F1c3   &  74.7(0.5)  &   10.1(1.1)  &   29(3)       &  &    0.47(0.09)  &    7(2)       &    19(6)       & $150 \pm 60$ &      3 \\
   F1c4   &  39.8(0.2)  &    5.4(0.6)  &   15.7(1.6)   &  &    0.14(0.04)  &    2.0(0.8)   &    6(2)        & $270 \pm 140$ &     4 \\
   F1c5   &   6.1(0.3)  &   0.8(0.12)  &   2.4(0.4)    &  &    $\leq 0.13$ &    $\leq 1.8$ &    $\leq 5.2$  & $\geq 50$  &    5 \\
   F1c6   &  14.6(0.2)  &   2.0(0.2)   &  5.7(0.7)     &  &    $\leq 0.08$ &    $\leq 1.1$ &    $\leq 3.1$  & $\geq 190$ &     6 \\
   F2c1   &  17.7(0.2)  &   2.4(0.3)   &  7.0(0.8)     &  &    $\leq 0.06$ &    $\leq 0.8$ &    $\leq 2.4$  & $\geq 290$ &      7 \\
 F3c1/2   &  17.4(0.1)  &   2.3(0.3)   &  6.8(0.7)     &  &    $\leq 0.06$ &    $\leq 0.8$ &    $\leq 2.5$  & $\geq 280$ &      8 \\
   F3c3   &  13.4(0.3)  &   1.8(0.2)   &  5.3(0.6)     &  &    $\leq 0.09$ &    $\leq 1.3$ &    $\leq 3.7$  & $\geq 140$ &      9 \\
   F3c4   &  14.1(0.2)  &   1.9(0.2)   &  5.6(0.6)     &  &    $\leq 0.07$ &    $\leq 1.0$ &    $\leq 2.9$  & $\geq 190$ &     10 \\
   F4c4   &  18.2(0.2)  &   2.5(0.3)   &   7.2(0.8)    &  &    $\leq 0.08$ &    $\leq 1.1$ &    $\leq 3.3$  & $\geq 220$ &     11 \\
   F4c5   &  25.8(0.2)  &   3.5(0.4)   &  10.2(1.1)    &  &    0.14(0.05)  &    2.0(0.8)   &    6(2)        & $175 \pm 90$ &     12 \\
   F4c6   &  29.1(0.3)  &   3.9(0.4)   &  11.4(1.3)    &  &    $\leq 0.11$ &    $\leq 1.5$ &    $\leq 4.4$  & $\geq 260$ &     13 \\
   F4c7   &  52.8(0.3)  &   7.1(0.8)   &  21(2)        &  &    0.27(0.7)   &    3.8(1.4)   &    11(4)       & $190 \pm 90$ &     14 \\
   F4c8   &  55.3(0.3)  &   7.4(0.8)   &  22(2)        &  &    $\leq 0.11$ &    $\leq 1.6$ &    $\leq 4.6$  & $\geq 470$ &     15 \\
   F4c9   &  35.0(0.3)  &   4.7(0.5)   &  13.8(1.5)    &  &    $\leq 0.11$ &    $\leq 1.6$ &    $\leq 4.6$  & $\geq 300$ &     16 \\
  F4c10   &  17.6(0.4)  &   2.4(0.3)   &  6.9(0.8)     &  &    $\leq 0.11$ &    $\leq 1.5$ &    $\leq 4.5$  & $\geq 160$ &     17 \\
   F4c1   &  13.0(0.2)  &   1.7(0.2)   &  5.1(0.6)     &  &    $\leq 0.14$ &    $\leq 2.0$ &    $\leq 5.9$  & $\geq 90$ &     18 \\
   F2c2   &  8.0(0.2)  &   1.1(0.13)   &  3.2(0.4)     &  &   $\leq 0.09$  &    $\leq 1.3$  &   $\leq 3.8$  & $\geq 84$ &    19  \\
   F3c5   &  19.4(0.3)  &   2.6(0.3)   &  7.6(0.8)     &  &   $\leq 0.09$  &    $\leq 1.3$  &   $\leq 3.8$  & $\geq 200$ &   20  \\
   \hline
   H3c4   &  15.4(0.2)  &   2.1(0.2)   &  6.1(0.7)     &  &    $\leq 0.10$ &    $\leq 1.3$ &    $\leq 3.9$  & $\geq 160$ &      1 \\
   H3c5   &  12.6(0.3)  &   1.7(0.2)   &  5.0(0.6)     &  &    $\leq 0.16$ &    $\leq 2.3$ &    $\leq 6.6$  & $\geq 75$ &      2 \\
   H3c6   &   8.6(0.3)  &   1.2(0.15)  &   3.4(0.4)    &  &    $\leq 0.17$ &    $\leq 2.4$ &    $\leq 7.1$  & $\geq 50$ &      3 \\
   H3c3   &  20.7(0.3)  &   2.8(0.3)   &  8.1(0.9)     &  &    0.12(0.03)  &    1.7(0.6)   &    5.1(1.8)    & $160 \pm 70$ &      4 \\
   H2c1   &  13.6(0.3)  &   1.8(0.2)   &  5.3(0.6)     &  &    $\leq 0.14$ &    $\leq 2.0$ &    $\leq 5.8$  & $\geq 90$ &      5 \\
   H3c2   &  24.0(0.2)  &   3.2(0.4)   &  9.5(1.0)     &  &    $\leq 0.11$ &    $\leq 1.6$ &    $\leq 4.6$  & $\geq 200$ &     6 \\
   H4c1   &  21.3(0.2)  &   2.9(0.3)   &  8.4(0.9)     &  &    $\leq 0.14$ &    $\leq 1.9$ &    $\leq 5.7$  & $\geq 150$ &     7 \\
   H3c1   &  17.2(0.3)  &   2.3(0.3)   &  6.8(0.7)     &  &    $\leq 0.14$ &    $\leq 2.0$ &    $\leq 5.8$  & $\geq 120$ &     8 \\
   H5c3   &  26.1(0.2)  &   3.5(0.4)   &  10.3(1.1)    &  &    0.31(0.06)  &    4.3(1.3)   &    13(4)       & $80 \pm 30$ &     9 \\
   H5c1   &  18.1(0.3)  &   2.4(0.3)   &  7.1(0.8)     &  &    $\leq 0.15$  &    $\leq 2.1$ &    $\leq 6.1$  & $\geq 120$ &    10 \\
   \hline
\end{longtable}
\twocolumn

\begin{table*}
\begin{center}
\caption{Same as Table~\ref{tab:columns} for the \H\ cores (Fig.~\ref{fig:dendro}). The ID letters in Col.~10
are used to identify the cores in Fig.~\ref{fig:dendro}. In Col.~11 we list if the core can be associated with 
one of the \D\ cores identified in Kong et al.~(\citeyear{kong17}). For $N_{\rm tot}$(\N15),
we used the area obtained from the Gaussian fit to the main hyperfine component for all sources except for C--b 
and F--l, for which we used the total integrated intensity of all hyperfine components detected.
}
\begin{tabular}{c c c c c c c c c c c}
\hline
source &  \multicolumn{3}{c}{\H}   & &    \multicolumn{3}{c}{\N15} & &  & \\
\cline{2-4} \cline{6-8} \\
            &  $\int T_{\rm B}d{\rm v}$ & \multicolumn{2}{c}{$N_{\rm tot}$} & & $\int T_{\rm B}d{\rm v}$ & \multicolumn{2}{c}{$N_{\rm tot}$} & \r1415 $^{(a)}$ & ID  & \D\ \\
            &    K \kms\    &   $\times 10^{13}$\cmq\     &   $\times 10^{13}$\cmq\     & & K \kms\       &   $\times 10^{11}$\cmq\     &   $\times 10^{11}$\cmq\    &   &  &   \\
            &                    &     $T_{\rm ex}$ = 10~K             &       $T_{\rm ex}$ = 50~K       &          &       & $T_{\rm ex}$ = 10~K & $T_{\rm ex}$ = 50~K   &   &   &  \\
\hline
  C1-n2hp-1  &  1.8(0.1)    &   2.2(0.3)    &  6.4(1.0)   &   &   $\leq 0.03$  &   $\leq 0.8$  &   $\leq 2.2$  &   $\geq 290$  & a & C1N \\
  C2-n2hp-1  &  4.36(0.02)  &   5.3(0.6)    &  15.0(1.6)  &   &   0.10(0.03)   &   1.5(0.6)    &   4.5(1.7)    &   350(190) & b  & C2C \\
  C2-n2hp-2  &  3.87(0.02)  &   4.7(0.5)    &  13.7(1.4)  &   &   0.10(0.02)   &   2.5(0.8)    &   7(2)      &   190(80) & c & C2A/B \\
  C5-n2hp-1  &  1.61(0.01)  &   2.0(0.2)    &  5.7(0.6)   &   &   $\leq 0.04$  &   $\leq 1.0$  &   $\leq 3.0$  &   $\geq 194$ & d & C5A \\
  C6-n2hp-1  &  1.82(0.02)  &   2.2(0.3)    &  6.4(0.7)   &   &   $\leq 0.07$  &   $\leq 1.8$  &   $\leq 5.2$  &   $\geq 125$ & e & \\
  \hline
  D3-n2hp-1  &  1.14(0.01)  &   1.4(0.2)    &  4.0(0.4)   &   &   $\leq 0.05$  &   $\leq 1.3$  &   $\leq 3.7$  &   $\geq 110$  & a & \\
  D6-n2hp-1  &  1.10(0.01)  &   1.3(0.2)    &  3.9(0.4)   &   &   $\leq 0.05$  &   $\leq 1.3$  &   $\leq 3.7$  &   $\geq 106$  & b & \\
  D5-n2hp-1  &  1.33(0.01)  &   1.6(0.2)    &  4.0(0.4)   &   &   0.05(0.01)   &   1.3(0.4)    &   3.7(1.1)    &   130(60)  & c & \\
  D5-n2hp-2 &  1.17(0.01)  &   1.4(0.2)    &  4.1(0.5)   &   &   $\leq 0.07$  &   $\leq 1.8$  &   $\leq 5.2$  &   $\geq 80$ & d & \\
  D7-n2hp-1  &  3.05(0.02)  &   3.7(0.4)    &  10.8(1.2)   &   &   0.05(0.01)   &   1.3(0.4)    &   3.7(1.1)    &   290(120)  & e & \\
  D5-n2hp-3  &  1.74(0.02)  &   2.1(0.3)    &  6.2(0.7)   &   &   $\leq 0.07$  &   $\leq 1.8$  &   $\leq 5.2$  &   $\geq 120$ & f & \\
  D9-n2hp-1  &  0.86(0.01)  &   1.0(0.1)    &  3.0(0.3)   &   &   $\leq 0.05$  &   $\leq 1.3$  &   $\leq 3.7$  &   $\geq 83$  & g & \\
  \hline
  F1-n2hp-1  &  4.55(0.01)  &   5.5(0.6)    &  16.1(1.6)  &   &   0.13(0.01)   &   3.3(0.6)    &   9.6(1.7)    &   170(50)  & a & \\
  F1-n2hp-2  &  0.45(0.01)  &   0.55(0.07)  &  1.6(0.2)   &   &   $\leq 0.05$  &   $\leq 1.3$  &   $\leq 3.7$  &   $\geq 43$  & b & F1 \\
  F1-n2hp-3  &  0.54(0.01)  &   0.65(0.08)  &  1.9(0.2)   &   &   $\leq 0.07$  &   $\leq 1.8$  &   $\leq 5.2$  &   $\geq 37$ & c & \\
  F1-n2hp-4  &  1.07(0.01)  &   1.3(0.2)    &  3.8(0.4)   &   &   $\leq 0.06$  &   $\leq 1.5$  &   $\leq 4.4$  &   $\geq 86$  & d & \\
  F1-n2hp-5  &  0.80(0.01)  &   1.0(0.1)    &  2.8(0.3)   &   &   $\leq 0.05$  &   $\leq 1.3$  &   $\leq 3.7$  &   $\geq 77$ & e & \\
  F2-n2hp-1  &  0.60(0.01)  &   0.73(0.08)  &  2.1(0.3)   &   &   $\leq 0.07$  &   $\leq 1.8$  &   $\leq 5.2$  &   $\geq 41$  & f & \\
  F2-n2hp-2  &  1.26(0.01)  &   1.5(0.2)   &  4.5(0.5)   &   &   $\leq 0.04$  &   $\leq 1.0$  &   $\leq 3.0$  &   $\geq 152$ & g & F2 \\
  F2-n2hp-3  &  0.93(0.02)  &   1.1(0.2)    &  3.3(0.4)   &   &   $\leq 0.07$  &   $\leq 1.8$  &   $\leq 5.2$  &   $\geq 64$  & h & \\
  F3-n2hp-1  &  1.88(0.01)  &   2.3(0.3)    &  6.7(0.7)   &   &   $\leq 0.04$  &   $\leq 1.0$  &   $\leq 3.0$  &   $\geq 227$ & i & \\
  F4-n2hp-1  &  2.52(0.01)  &   3.1(0.4)    &  8.9(0.9)   &   &   0.056(0.008) &   0.8(0.2)    &   2.3(0.6)    &   390(140) & l & \\
  F4-n2hp-2  &  0.48(0.01)  &   0.58(0.07)  &  1.7(0.2)   &   &   $\leq 0.06$  &   $\leq 1.5$  &   $\leq 4.4$  &   $\geq 39$ & m & \\
  F4-n2hp-3  &  0.98(0.02)  &   1.2(0.2)    &  3.5(0.4)   &   &   $\leq 0.04$ &   $\leq 1.0$    &   $\leq 3.0$    &   $\geq 120$  & n & \\  
  \hline
  H1-n2hp-1  &  1.04(0.01)  &   1.3(0.2)    &  3.7(0.4)   &   &   $\leq 0.03$  &   $\leq 0.8$  &   $\leq 2.2$  &   $\geq 167$  & a & \\
  H1-n2hp-2  &  1.61(0.01)  &   2.0(0.2)    &  5.7(0.6)   &   &   $\leq 0.03$  &   $\leq 0.8$  &   $\leq 2.2$  &   $\geq 260$  & b & H1A \\
  H3-n2hp-1  &  1.24(0.01)  &   1.5(0.2)   &  4.4(0.5)   &   &   $\leq 0.02$  &   $\leq 0.5$  &   $\leq 1.5$  &   $\geq 300$ & c & \\
  H3-n2hp-2  &  1.04(0.01)  &   1.3(0.2)    &  3.7(0.4)   &   &   $\leq 0.02$  &   $\leq 0.5$  &   $\leq 1.5$  &   $\geq 251$ & d & H3A \\
  H2-n2hp-1  &  1.43(0.01)  &   1.7(0.2)   &  5.1(0.6)   &   &   0.027(0.008) &   0.7(0.3)    &   2.0(0.8)    &   260(130) & e & H2A/B/C/D/E/G \\
  H5-n2hp-1  &  1.41(0.01)  &   1.7(0.2)   &  5.0(0.5)   &   &   0.05(0.012)  &   1.3(0.4)    &   3.7(1.3)    &   140(60)  & f & H5A \\
\hline
\end{tabular}
\label{tab:columns-letters}
\end{center}
$^{(a)}$ derived from the column densities estimated assuming \Tex = 10~K, but the ratios are identical when using \Tex = 50~K.
\end{table*}

\section{Discussion}
\label{discu}

\subsection{\r1415\ ratios in the continuum cores}
\label{1415-cont}

In Fig.~\ref{fig:ratios} we plot $N_{\rm tot}$(\N15) against $N_{\rm tot}$(\H) computed for the 
continuum cores. Their ratios, \r1415\ (shown in Table~\ref{tab:columns}), are in between $\sim 80$ 
and $\sim 400$, with little statistically significant differences among the clouds: the average \r1415\ is 
$230\pm 50$ for Cloud C, $180\pm 100$ for Cloud D, 260$\pm 100$ for Cloud F, and 120$\pm 40$ 
for Cloud H. The cores have also been divided in Fig.~\ref{fig:ratios} in two groups: starless or star-forming, 
based on the absence or presence, respectively, of an embedded infrared emission. This classification
was made in paper I, and we will discuss the \r1415\ in the two groups in Sect.~\ref{activity}.

Inspection of Fig.~\ref{fig:ratios} indicates: 
\begin{itemize}
\item (1) a variability in the \r1415\ ratio of a factor $\sim 4$, smaller than the factor $\sim 10$ 
found in other IRDCs and high-mass star-forming regions at lower angular 
resolution by Zeng et al.~(\citeyear{zeng17}) and Fontani et al.~(\citeyear{fontani15}), but
similar to that measured in HCN and HNC by Colzi et al.~(\citeyear{colzi18b}); 
\item (2) an average \r1415\ of $\sim 210$, very similar to the terrestrial one
(Marty et al.~\citeyear{marty09}). Assuming Galactocentric distances for the four 
clouds in between $\sim 4.6$ and $\sim 6.3$~kpc (Zeng et al.~\citeyear{zeng17}), the 
present-day \r1415\ ratio of the ISM at the Galactocentric distance of the sources is $\sim 300-330$, 
estimated from the trend derived by Colzi et al.~(\citeyear{colzi19}, see their Eqs. 5 and 6) 
from HCN and HNC isotopologues. Hence, the average \r1415\ ratio measured in this work is 
$\sim 1.5$ times smaller than the ISM value at the Galactocentric distance of the sources;
\item (3) in two cores, i.e.~D6c4 and H5c3, we measure \r1415\ ratios around 100, i.e. lower by a 
factor $\sim 3$ with respect to the local ISM value, and comparable to (or slightly smaller than) 
the cometary values. 
\end{itemize}

In Fig.~\ref{fig:ratios} we highlight the difference with the single-dish survey of Fontani et 
al.~(\citeyear{fontani15}), in which the measured range of \r1415\ from \H\ isotopologues is 
$\sim 180 - 1300$, namely twice wider than ours and hence with a higher average value. 
A wider range in \r1415\ was also measured in HCN and HNC towards the IRDC cores 
studied by Zeng et al.~(\citeyear{zeng17}) with a single-dish.
Also, the two cores D6c4 and H5c3 have among the lowest \r1415\ ratios ever measured in \H,
and similar to the values obtained by Colzi et al.~(\citeyear{colzi19}) towards
IRAS 05358+3543. This would indicate that the smaller spatial scales probed by the ALMA 
interferometer can identify spots with lower \r1415\ values much better than single-dish
observations can. 
 
To investigate more in detail the influence of the diffuse gas surrounding the dense cores on the 
measured \r1415, we have derived the \r1415\ ratios towards the cores detected in \N15\ 
from the total power data only. We have extracted and analysed the total power spectra following 
the same approach described in Sects.~\ref{detection-cont} and \ref{coldens-cont}. The only difference is 
that in Eq.~\ref{eq:conversion} we used the telescope full width at half maximum to convert from
flux density to main beam temperature units. The half power beam width of the total power data is 
$\sim 70$\asec\ for both \N15\ and \H\ (1--0), so that the spectra extracted from
one position can contain the contribution of nearby cores. Therefore, in case of cores 
separated by less than half of the beam width, i.e. $\sim 35$\asec, we have extracted
only one single-dish spectrum. The results are shown in Table~\ref{tab:single-dish}. In 
all cores except F1c1, which is the most extended one, the beam-averaged \r1415\ ratios 
are larger by a factor $\sim 1.5 - 3$ than those obtained in the embedded cores at higher angular 
resolution. This result strengthens our conclusion that in the diffuse envelope of the cores 
the isotopic ratio tends to be larger, as predicted by the selective photodissociation scenario,
and also stresses the need for high-angular resolution data to derive correct values of
the \r1415\ ratio and accurately measure the $^{15}$N fractionation in the cores embedded 
within the IRDC.

\begin{table*}
\begin{center}
\caption{Integrated line intensities ($\int T_{\rm MB}d{\rm v}$), total column densities ($N_{\rm tot}$)
of \H\ and \N15, and their ratio \r1415, obtained from the spectra extracted from the total power data only. 
The central position of the extracted spectra is the core listed in Col.~1 (see Sect.~\ref{res}), and the
extraction region corresponds to the total power beam width at half maximum ($\sim 70$\asec).
For \H(1--0), the listed $\int T_{\rm MB}d{\rm v}$ is that of the F$_1$=0--1 hyperfine component only,
which has a relative intensity of 0.111 with respect to the total one. For \N15(1--0) the listed 
$\int T_{\rm MB}d{\rm v}$ is that of the F=2--1 hyperfine component only, which has a relative 
intensity of 0.555 with respect to the total one. }
\begin{tabular}{c c c c c c c c c }
\hline
\hline
source &  \multicolumn{3}{c}{\H}   & &    \multicolumn{3}{c}{\N15} & \\
\cline{2-4} \cline{6-8} \\  
       & $\int T_{\rm MB}d{\rm v}$ & \multicolumn{2}{c}{$N_{\rm tot}$} & & $\int T_{\rm MB}d{\rm v}$ & \multicolumn{2}{c}{$N_{\rm tot}$}  & \r1415\ \\
            &    K \kms\    &   $\times 10^{13}$\cmq\     &   $\times 10^{13}$\cmq\     & & K \kms\       &   $\times 10^{11}$\cmq\     &   $\times 10^{11}$\cmq\ &    \\
            &                    &     $T_{\rm ex}$ = 10~K             &       $T_{\rm ex}$ = 50~K       &          &       & $T_{\rm ex}$ = 10~K & $T_{\rm ex}$ = 50~K   &    \\
\hline
C2c4   & 34.5(0.2)             &     4.2(0.4)  & 12.2(1.3)                           &  & 0.33(0.03)             &   8.3(1.6)  & 24(5) & 500(150) \\
D6c4   & 26.7(0.2)             &     3.2(0.3)  &   9.5(1.0)                           &  & 0.35(0.03)             &   8.8(1.6)  & 26(5) & 370(100) \\
D8c1     & 30.5(0.2)             &     3.7(0.4)  & 10.8(1.2)                           &  & 0.45(0.05)             &  11(2)       & 33(7)  & 330(100) \\
F1c1       & 67.9(0.2)             &     8.2(0.8)  &   24(3)                               &  & 0.78(0.09)             &   20(4)      & 60(12) & 420(130) \\
F4c7     & 21.8(0.2)             &     2.6(0.3)  &   7.7(0.8)                           &  & 0.27(0.04)             &   6.8(1.7)  & 20(5)  & 390(140) \\
H3c3       & 16.8(0.2)             &     2.0(0.2)  &   6.0(0.7)                           &  & 0.28(0.04)             &  7.0(1.7)   &   21(5) & 290(100) \\
H5c3       & 14.0(0.2)             &     1.7(0.2)  &   5.0(0.6)                           &  & 0.24(0.04)             &  6.0(1.6)   &   18(5) & 280(100) \\
\hline
\end{tabular}
\label{tab:single-dish}
\end{center}
\end{table*}

However, as one can see in Fig.~\ref{fig:ratios}, all lower limits on \r1415\ are below $\sim 400$,
and the maximum measured \r1415\ ratios are $\sim 390$ and $\sim 400$ towards F1c1 and F2c2,
respectively. This means that our interferometric observations are not sensitive to \N15\ emission 
in cores with \r1415\ higher than $\sim 400$, and hence we cannot conclude if the narrower 
range of isotopic ratios measured in this work depends also on our limited sensitivity.
Another caveat on the \r1415\ ratios arises from the possible non negligible optical depth of
the F$_1=0-1$ component of \H(1--0), which could lead us to underestimate $N_{\rm tot}$(\H) and,
from this, the \r1415\ ratio. This comment applies especially to core H5c3, which has the lowest \r1415\
ratio ($\sim 80$). The fit to the hyperfine structure of this line provides an optical depth of the main
component $\tau \sim 8$, and hence $\tau$ of the F$_1=0-1$ component would be $\sim 0.89$
(Fig.~\ref{fig:spectra-cloudH}).
This would introduce a correction $\tau/(1-{\rm exp}(-\tau))\sim 1.7$ in the \H\ column density, and 
hence the \r1415\ would become $\sim 140$, still about a factor 2 smaller than the ISM value
at the Galactocentric distance of the source.

\subsection{\r1415\ ratios in the \H\ cores}
\label{1415-n2hp}

\begin{figure}
\centering
{\includegraphics[width=8cm,angle=0]{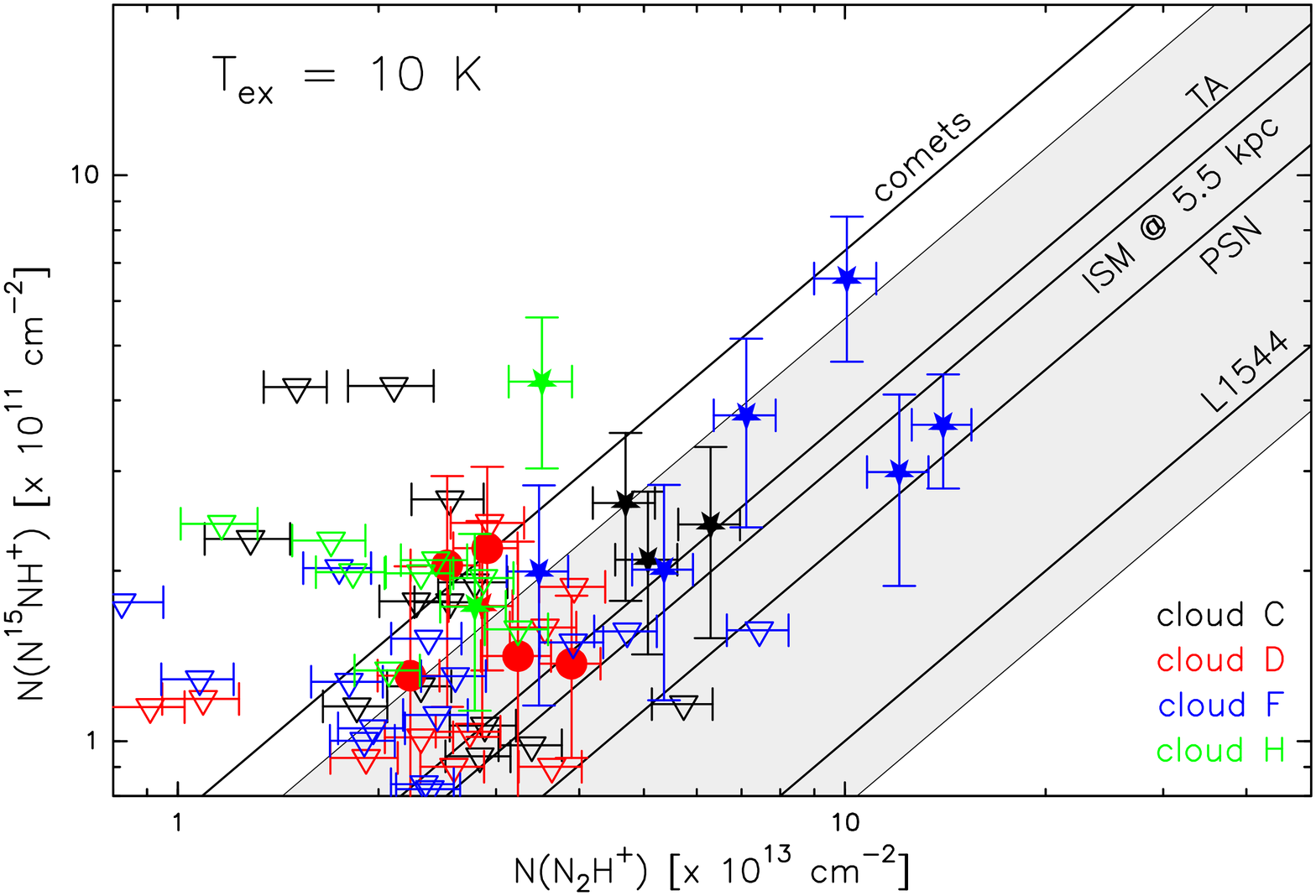}
\includegraphics[width=8cm,angle=0]{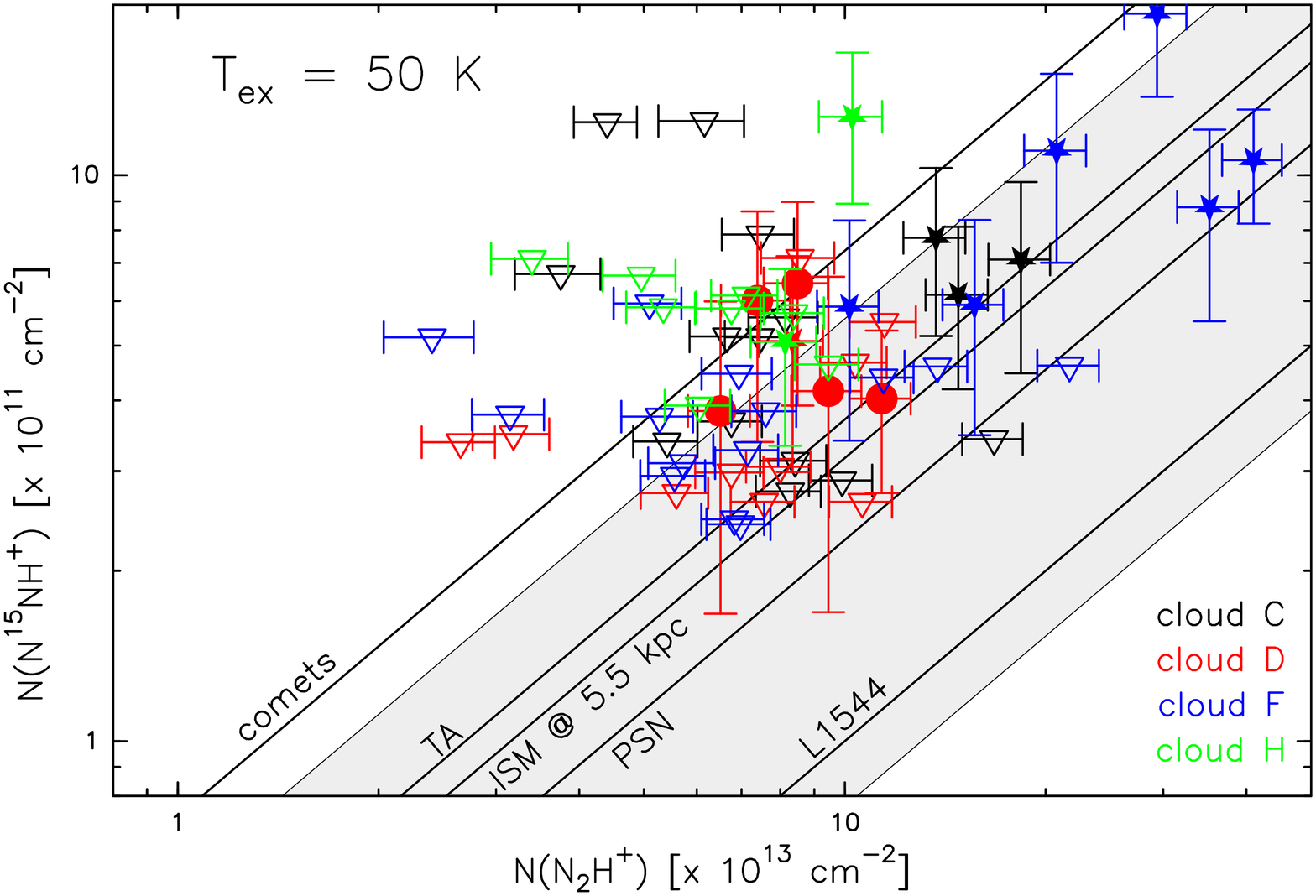}
}
\caption{Column density of \H\ against that of \N15\ assuming \Tex = 10~K (top panel) and
\Tex = 50~K (bottom panel). The values refer to the spectra extracted from the continuum
cores (see Fig.~\ref{fig:dendro}).
In both panels, the filled symbols represent the sources detected in \N15, and the open triangles correspond 
to the upper limits on $N$(\N15 ). Among the detected sources, stars indicate cores with signs of star 
formation (in the form of embedded infrared sources), while circles correspond to cores without protostellar 
activity. The colours indicate the cores extracted from the different clouds
as labelled in the bottom-left corner. The solid lines indicate the mean \r1415\ composition as measured
in: comets ($\sim 136$, Shinnaka et al.~\citeyear{shinnaka16}); the terrestrial atmosphere
(TA, $\sim 270$, Marty et al.~\citeyear{marty09}); the ISM at the average Galactocentric
distance of the sources of 5.5~kpc ($\sim 316$, average value derived from the trends of 
Colzi et al.~\citeyear{colzi18b});
the protosolar nebula (PSN, $\sim 441$, Marty et al.~\citeyear{marty10}); the pre-stellar core 
L1544 ($\sim 1000$, Bizzocchi et al.~\citeyear{bizzocchi13}). 
The grey band indicates the range of \r1415\ measured by Fontani et al.~(\citeyear{fontani15}) in
a sample of high-mass star-forming cores with the IRAM-30m telescope.}
\label{fig:ratios}
\end{figure}

\begin{figure}
\centering
{\includegraphics[width=8cm,angle=0]{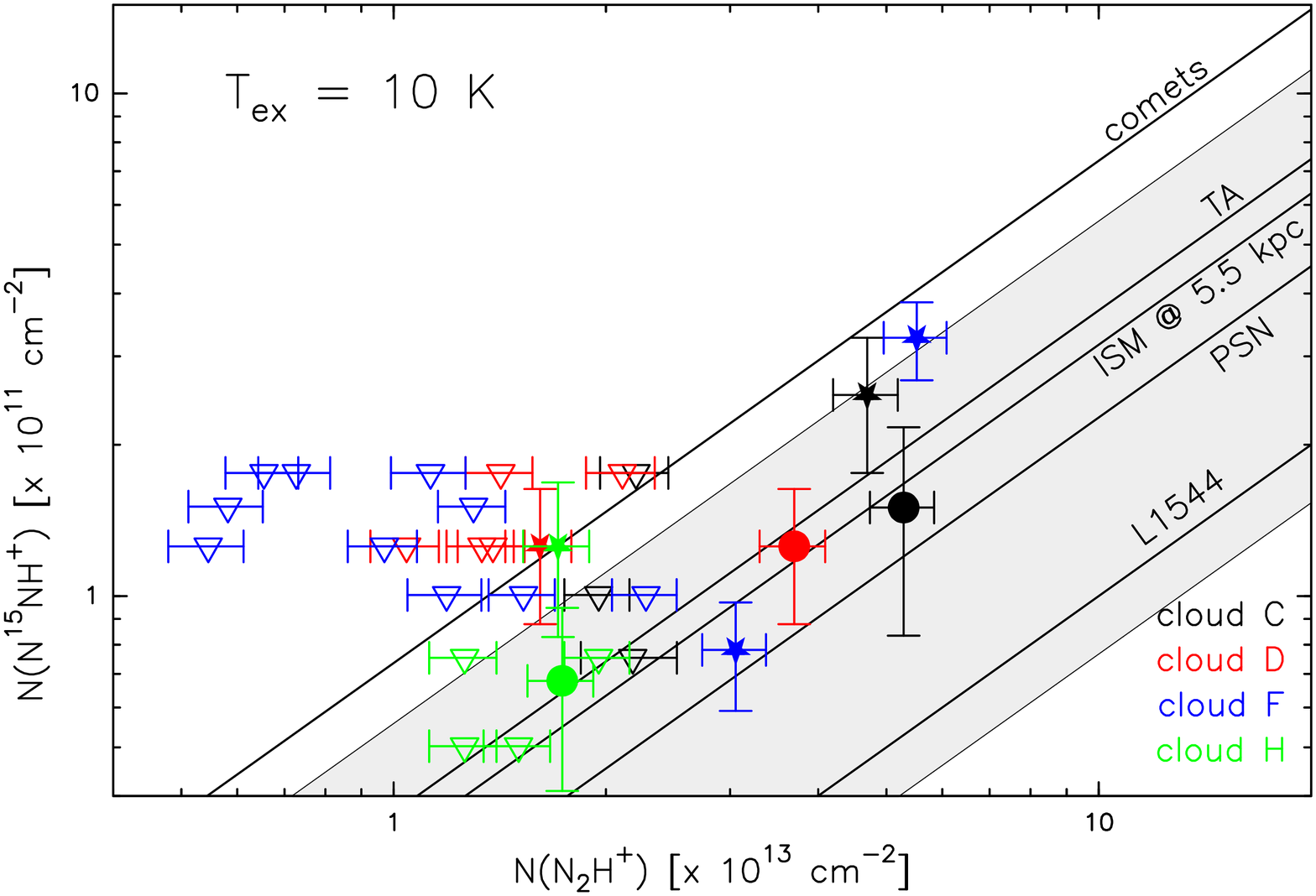}
\includegraphics[width=8cm,angle=0]{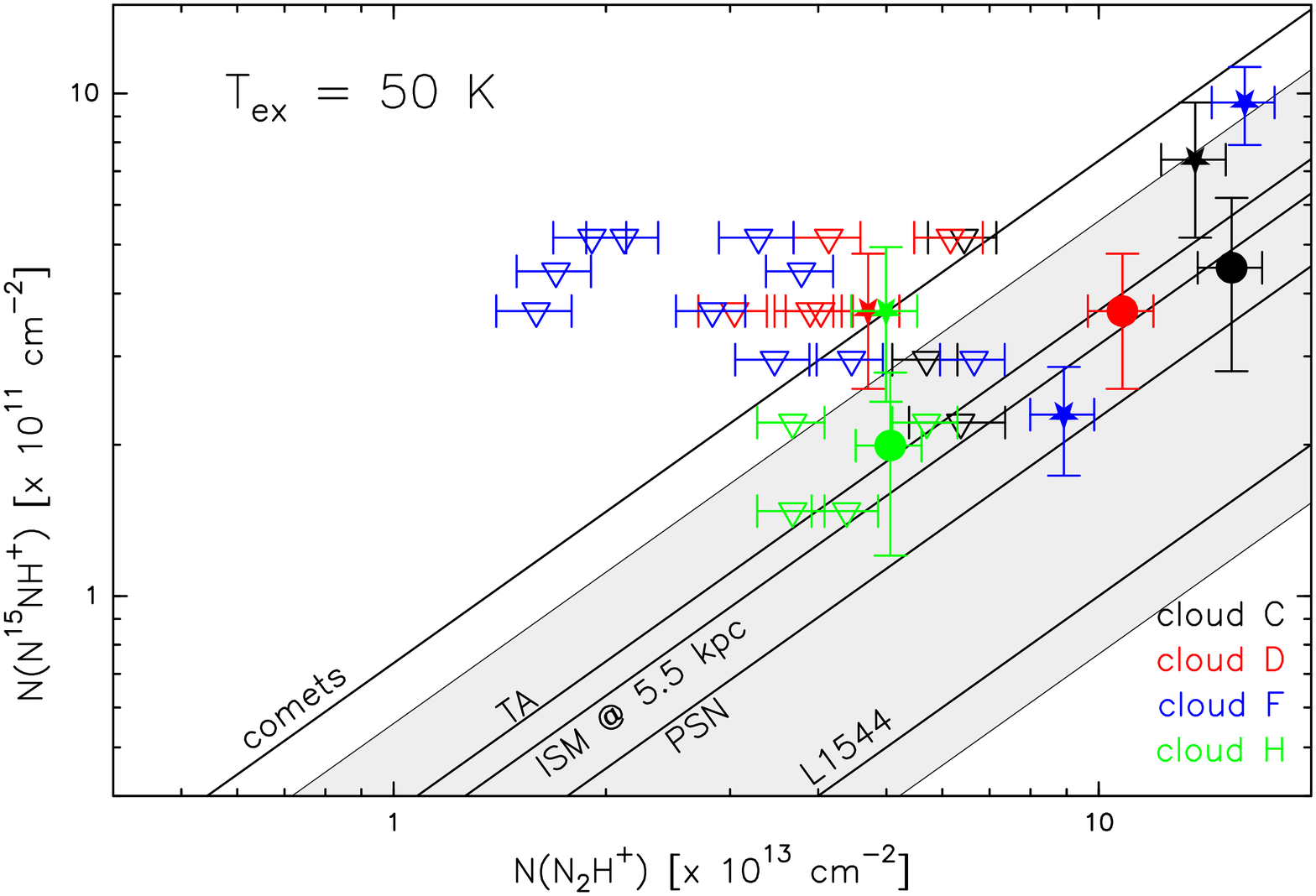}
}
\caption{Same as Fig.~\ref{fig:ratios} for the values derived from the spectra extracted from the
\H\ cores (see Fig.~\ref{fig:dendro}).}
\label{fig:ratios-letters}
\end{figure}

Let us now analyse the \r1415\ ratio towards the \H\ cores. The \H\ and \N15\ column densities
are listed in Table~\ref{tab:columns-letters}, and shown in Fig.~\ref{fig:ratios-letters}: the range of 
\r1415\ is $\sim 130 - 390$, consistent with the one found in the continuum cores 
(compare Fig.~\ref{fig:ratios} and Fig.~\ref{fig:ratios-letters}). For the \H\ cores encompassing 
continuum cores, all \r1415\ are consistent among them within the uncertainties, or slightly higher
(about a factor $\sim 1.5-2$, like in C--b, F--l and H--f), which indicates that the \r1415\ ratio 
in the core envelopes is either similar to or higher than that measured in their embedded dense
continuum cores. 

Overall, the non-detection of \N15\ in \H\ cores without continuum cores, and the fact that the \r1415\ ratios in 
the \H\ cores is (almost) never smaller than the values measured in their embedded 3~mm continuum 
cores suggests that the \r1415\ ratio tends to be higher in the more extended and diffuse gas. This result 
is in agreement with the high-angular resolution study of Colzi et al.~(\citeyear{colzi19}) 
towards the high-mass protocluster IRAS 05358+3543: in this source, the authors
show that in the individual star-forming cores the \r1415\ ratio is lower by a factor two than in the
extended, diffuse envelope. Colzi et al.~(\citeyear{colzi19}) concluded that this could be
due to selective photodissociation (Heays et al.~\citeyear{heays14}, Lee et al.~\citeyear{lee21}). 
This mechanism would 
act like this: because UV photons cause photodissociation of $^{14}$N$^{15}$N, the progenitor 
of \N15, but not of the self-shielded $^{14}$N$_2$, the progenitor of \H, one would expect a 
decrease of the \N15\ abundance with respect to \H\ in the external layers exposed to external
interstellar UV photons (and hence an increase of the \r1415\ ratio).
However, care needs to be taken in this interpretation because most upper limits are comparable to 
the measured \r1415\ in the detected cores, and hence to show that the \r1415\ ratio is higher
in the external, more diffuse part of the dense cores, higher sensitivity measurements of the \N15\ column
densities are needed. 

An interesting exception is F1-n2hp-1: its \r1415\ ratio is $170\pm 50$, while those of the four 3~mm 
continuum cores embedded in it vary from $\sim 390$ to $\sim 150$. Only F1c3 has a \r1415\ 
very similar to that of F1-n2hp-1 (150 against 170, respectively), while the others, F1c1, F1c2, and F1c4, 
have ratios higher by a factor $\sim 1.5 - 2$.
Hence, in these cases the \r1415\ ratio associated with the diffuse envelope of the continuum
cores has a lower \r1415, at odds with selective photodissociation if the UV photons are produced 
by an external field. A possible explanation still consistent with 
the selective photodissociation scenario is that internal UV photons are produced by the
protostar(s) embedded in F1c1 instead of by an external UV field, and hence in this case the \N15\
abundance is expected to be lower (and hence the \r1415\ ratio to be higher) in the dense
protostellar core than in the envelope. Also, F1-n2hp-1 is very extended (Fig.~\ref{fig:dendro}) and 
in principle can include gas not only associated with the external envelope of F1c1, F2c2, F2c3, 
and F2c4, but also with other gaseous components having possibly different (and unknown) 
physical properties.

%

\subsection{Relation between \r1415\ and star-formation activity}
\label{activity}

As stated in Sect.~\ref{1415-cont}, in paper I the 3~mm continuum cores have been classified 
as starless or star-forming based on the absence or presence, respectively, of infrared emission. 
We can check if the measured \r1415\ ratios changes with the presence/absence of star-formation 
activity within the cores. In Fig.~\ref{fig:ratios}, we indicate with different symbols the starless 
and star-forming cores. Among the 13 detections, we find five starless and 12 star-forming cores, 
respectively. The average \r1415\ ratio is slightly lower ($\sim 190$) in the starless cores than 
in the protostellar objects ($\sim 220$). However, given the large dispersions, we confirm overall 
the result suggested by previous works (Fontani et al.~\citeyear{fontani15}, Zeng et al.~\citeyear{zeng17}, 
Colzi et al.~\citeyear{colzi18a},~\citeyear{colzi18b}): time does not seem to play a role in the 
enrichment, or depletion, of $^{15}$N in \H.

Interestingly, the \r1415\ ratio in starless cores seems slightly lower on average than the
values typically measured towards well known pre-stellar cores (e.g., by Bizzocchi et 
al.~\citeyear{bizzocchi13}, Redaelli et al.~\citeyear{redaelli18}). However, again care needs to
be taken in the interpretation of this result. In fact, first the nature of our starless cores could not
be "pre--stellar" but just starless or star-forming in a very early stage of star-formation activity;
second, some cores are extended and could contain smaller fragments in different evolutionary 
stages. For example, a core considered as starless could encompass smaller fragments already 
star-forming undetected in the IR images (see e.g.~Liu et al.~\citeyear{liu18}). Therefore,
in summary the results confirm the previous claim that evolution does not seem to play
a role in the fractionation of nitrogen, but observations at higher sensitivity and angular
resolution are needed to confirm or deny this conclusion.

\subsection{Relation between \r1415\ and core properties}
\label{comparison}

\begin{figure}
\centering
{\includegraphics[width=8cm,angle=0]{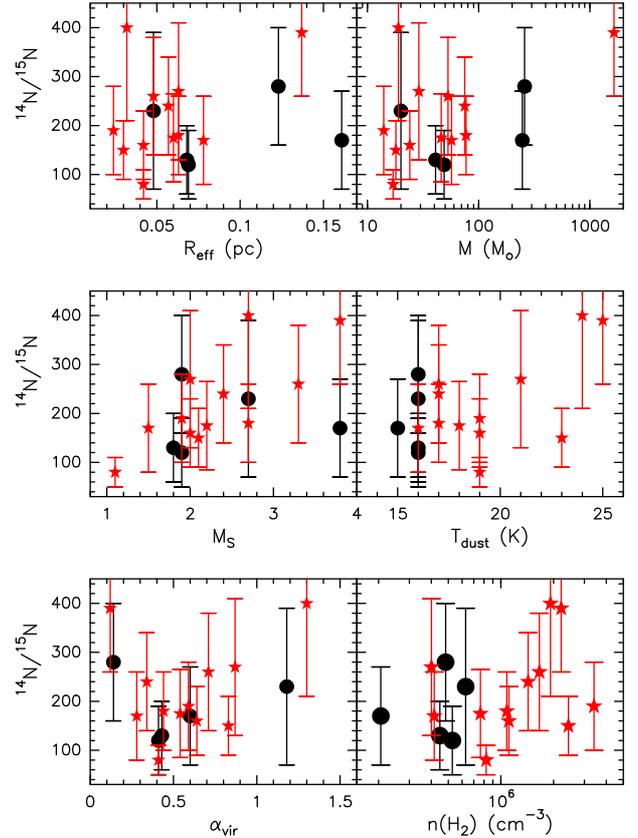}
}
\caption{\r1415\ ratio of the continuum cores detected in \N15\ as a function of core physical 
properties derived in paper I. Specifically, we plot the \r1415\ versus: radius (top-left), mass (top-right), Mach 
number (centre-left), dust temperature (centre-right), virial parameter (bottom-left) and H$_2$ volume 
density (bottom-right). In each frame, red stars indicate cores with embedded star-formation activity,
and black circles the starless cores.}
\label{fig:properties}
\end{figure}

\begin{figure}
\centering
{\includegraphics[width=8cm,angle=0]{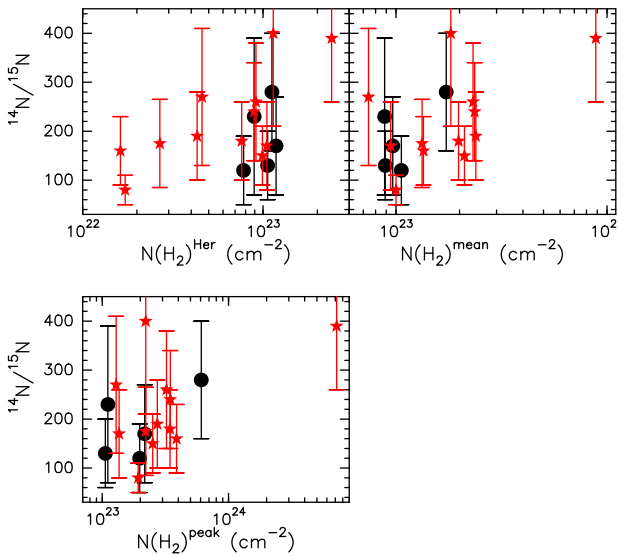}
}
\caption{Same as Fig.~\ref{fig:properties} for the H$_2$ column densities of the cores calculated
from the Herschel continuum (top-left) and from the mean and peak 3~mm continuum emission
(top-right and bottom-left, respectively).}
\label{fig:properties-2}
\end{figure}

In Figs.~\ref{fig:properties} and ~\ref{fig:properties-2}, we compare the \r1415\ ratios in the 17 
continuum cores detected in \N15\ with some physical properties that were derived in paper I, 
namely: radius ($R_{\rm eff}$), gas mass ($M$), dust temperature ($T_{\rm dust}$, used to 
compute $M$), sonic Mach number ($\mathcal{M}_{\rm S}$), virial parameter ($\alpha_{\rm vir}$), 
H$_2$ volume density ($n{\rm (H_2)}$), and H$_2$ column density ($N{\rm (H_2)}$). This latter 
was derived in three ways: from the Herschel data ($N{\rm (H_2)}^{\rm Her}$), from the 3~mm 
continuum mean flux density ($N{\rm (H_2)}^{\rm mean}$) and from the 3~mm continuum 
peak flux density ($N{\rm (H_2)}^{\rm peak}$). 
For quantities that in paper I have been derived from background subtracted and non-subtracted
data, such as the mass and the H$_2$ volume density, we have adopted the non-subtracted data.
The parameters derived from the Herschel maps (i.e.~$T_{\rm dust}$ and $N{\rm (H_2)}^{\rm Her}$)
were obtained by resampling the Herschel maps to the higher resolution of the ALMA data via a 
bilinear interpolation. The methods used to derive each parameter are described in paper I. 

\begin{table*}
\begin{center}
\caption{Core properties derived in paper I: radius ($R_{\rm eff}$), mass ($M$), dust temperature ($T_{\rm dust}$,
used to compute $M$), sonic Mach number ($\mathcal{M}_{\rm S}$), virial parameter ($\alpha_{\rm vir}$), 
H$_2$ volume density ($n{\rm (H_2)}$), and H$_2$ column density derived in three ways: from the 
Herschel data ($N{\rm (H_2)}^{\rm Her}$), from the 3~mm mean continuum flux density ($N{\rm (H_2)}^{\rm mean}$)
and from the 3~mm peak flux density ($N{\rm (H_2)}^{\rm peak}$). $M$, $\alpha_{\rm vir}$, $n{\rm (H_2)}$, 
and $N{\rm (H_2)}$) were calculated from the background subtracted 3~mm flux densities.}
\begin{tabular}{c c c c c c c c c c c}
\hline
\hline
core  & \r1415\ $^{(a)}$  & $R_{\rm eff}$  &  $M$  &  $T_{\rm dust}$ &   $\mathcal{M}_{\rm S}$ &  $\alpha_{\rm vir}$ &  $n$(${\rm H_2}$)    &  $N$(${\rm H_2}$)$^{\rm Her}$   & $N$(${\rm H_2}$)$^{\rm mean}$  &  $N$(${\rm H_2}$)$^{\rm peak}$  \\
         &                            &  pc    & M$_{\odot}$ &  K       &            &                &   $\times 10^5$\cmc\                  &  \multicolumn{3}{c}{$\times 10^{23}$cm$^{-2}$} \\
         \hline
\multicolumn{11}{c}{starless cores} \\
\hline
D6c4 &  120 &   0.069   &  49   &    16   &   1.9   &   0.41   &  5.28   &   0.78   &   1.06   &  1.98  \\
D6c6 &  230 &    0.048  &   20  &     16  &    2.7  &    1.18  &   6.29  &   0.89  &    0.88  &   1.11  \\
D7c2 &  130  &   0.068   &  41   &    16   &   1.8   &   0.43   &  4.48   &   1.06   &   0.89   &  1.06  \\
D5c5/6 &  170  &    0.161  &  247  &     15  &    3.8  &    0.60  &   2.06  &   1.18  &    0.97  &   2.17  \\
D8c1 &  280  &    0.123  &  259  &     16  &    1.9  &    0.14  &   4.83  &   1.12  &    1.73  &   6.08  \\
\hline
\multicolumn{11}{c}{star-forming cores} \\
\hline
C2c3/5 &  240   &    0.057  &   75  &     17  &    2.4  &    0.34  &   14.3  &   0.89  &    2.36  &   3.46  \\
C2c4 &  260   &    0.048  &   53  &     17  &    3.3  &    0.71  &   16.6  &   0.91  &    2.32  &   3.23  \\
C2c6 &  180    &   0.063   &  77   &    17   &   2.7   &   0.44   &   10.8   &  0.76   &   1.98   &  3.44  \\
D7c1 &  170    &   0.078   &  57   &    16   &   1.5   &   0.28   &   4.16   &  1.05   &   0.94   &  1.36  \\
F1c1 &  390   &    0.137  &  1650 &      25 &     3.8 &     0.12 &   22.1  &   2.40  &     8.85  &  71.6  \\
F1c2 &  400   &    0.032  &   19  &     24  &    2.7  &    1.30  &   19.2  &   1.14  &    1.82   &  2.21  \\
F1c3 &  150    &   0.030   &  18   &    23   &   2.1   &   0.83   &   24.3   &  0.99   &   2.11   &  2.51  \\
F1c4 &  270   &    0.063  &   29  &     21  &    2.0  &    0.87  &   4.01  &   0.46   &   0.74   &  1.29  \\
F4c5 &  175    &   0.060   &  46   &    18   &   2.2   &   0.54   &  7.64   &   0.27   &   1.33   &  2.21  \\
F4c7 &  190    &   0.024   &  14   &    19   &   1.9   &   0.59   &  34.0   &   0.43   &   2.39   &  2.73  \\
H3c3 &  160   &   0.042   &  24   &    19   &   2.0   &   0.64   &  11.1   &   0.16   &   1.35   &  3.88  \\
H5c3 &  80     &   0.042   &  17   &    19   &   1.1   &   0.41   &  8.24   &   0.17   &   1.00   &  1.93  \\
\hline
\end{tabular}
\end{center}
$^{(a)}$ from Table~\ref{tab:columns};
\end{table*}

Inspection of Figs.~\ref{fig:properties} and~\ref{fig:properties-2} suggests that, overall, there are no 
strong and clear (anti-)correlations between the \r1415\ ratio and the analysed core properties. However, 
non-parametric statistical tests suggest very tentative positive correlations with the 
H$_2$ column density derived from the 3~mm continuum (in Fig.~\ref{fig:properties-2}).
The Pearson's $r$ correlation coefficient is: $r\sim 0.61$ with $N{\rm (H_2)^{Her}}$ ($p-$value 0.001); 
$r\sim 0.58$ with $N{\rm (H_2)^{mean}}$ ($p-$value 0.007); $r\sim 0.54$ with $N{\rm (H_2)^{peak}}$ 
($p-$value 0.01).

We speculate that the positive correlations between the \r1415\ ratios and the core column densities 
could again be the consequence of the selective  photodissociation mechanism, already invoked in 
Sect.~\ref{1415-n2hp}. In fact, the higher \r1415\ average ratio could be due to a larger amount of gas 
along the line of sight in the core envelope, more exposed to external UV photons responsible for the 
photodissociation of the $^{14}$N$^{15}$N molecule (the parent species of \N15) but not of the 
self-shielded $^{14}$N$_2$ (parent species of \H), thus causing an increase in the \r1415\ ratio.
If we analyse only the star-forming cores, however, we find a tentative positive correlation 
between \r1415\ and the sonic Mach number $\mathcal{M}_{\rm S}$ ($r\sim 0.76$, $p-$value 0.004).
A tentative difference in $\mathcal{M}_{\rm S}$, although not statistically significant, was found
in paper I between starless and star-forming cores, with the star-forming cores being associated
with larger $\mathcal{M}_{\rm S}$. A very tentative one can be also found between \r1415\ and 
$T_{\rm dust}$ ($r\sim 0.6$, $p-$value 0.04). Because both parameters
are usually associated with gas in a more advanced stage of star-formation activity, these tentative
correlations could indicate that more evolved cores may be associated with higher
\r1415\ ratios. 
However, all trends are strongly influenced by sources F1c1 and F1c2 (Figs.~\ref{fig:properties} and 
\ref{fig:properties-2}). Moreover, most \N15\ lines are detected with a signal to noise ratio in 
between 3 and 5, and hence any trend that can be found from the plots in Figs.~\ref{fig:properties} 
and~\ref{fig:properties-2} must be taken with caution.


Overall, the fact that we find either no or little correlation between \r1415\ and core properties 
confirm the chemical models of Roueff et al.~(\citeyear{roueff15}) and Loison et al.~(\citeyear{loison19}), 
in which enrichment in $^{15}$N is always very low or even negligible for all species 
formed in the gas phase like \H. In Loison et al.~(\citeyear{loison19}), a significant $^{15}$N enrichment 
is possible if the gas temperature is below $\sim 10$~K, due to a possibly different exothermicity of the 
formation reactions of \H\ and \N15, not yet experimentally measured (see also Hily-Blant et al.~\citeyear{hily-blant20}). 
However, the temperature of our cores at the scales studied by us is always above $\sim 10$~K, in 
agreement with the lack of $^{15}$N enrichment, although the temperature measurement is taken over
an angular scale (i.e. the Herschel beam) larger than that of the individual cores. Therefore, in quiescent cores
the temperature could be lower on smaller scales, as discussed in paper I.

\section{Conclusions}
\label{conc}

We have derived the \r1415\ ratio from the \H\ and \N15\ (1--0) lines in a sample of four well-studied IRDCs. 
The observations, performed with ALMA, represent the first interferometric survey of the \r1415\ ratio in cores 
embedded in IRDCs. The main results are summarised below:
\begin{itemize}
\item among the 3~mm continuum cores embedded in each IRDC and identified in paper I, 
we detect 17 out of 62 cores in \N15; 
\item the \r1415\ values derived from the column density ratio $N$(\H)/$N$(\N15) in all
studied regions are in the range $\sim 80 - 400$, which roughly goes from the cometary values
to the ISM value at the Galactocentric distance of the targets;
\item the \r1415\ ratios are smaller on average than those found in previous single-dish studies, 
in which ratios even exceeding $\sim 1000$ are found, suggesting that the \r1415\ ratio in the 
cores is smaller than that in their associated envelope, in agreement with the selective photodissociation scenario. 
The conclusion is strongly supported by the fact that the \r1415\ ratios obtained using the total power data only 
towards the detected cores are on average a factor 1.5--3 higher. This also highlights the need 
for high-angular resolution measurements to measure accurately the \r1415\ ratio in dense 
cores within IRDCs. However, the sensitivity of our observations likely does 
not allow us to derive \r1415\ ratios higher than $\sim 400$ at small scales. Therefore, 
higher sensitivity measurements will be absolutely required to confirm, or deny, this finding;
\item the average \r1415\ ratios in the four clouds (230, 180, 260, 120), are marginally lower
(by a factor $\sim 1.5-2$) than the present-day Galactic value at the Galactocentric distance 
of the clouds ($\sim 300-330$);
\item We find tentative positive trends between \r1415\ and H$_2$ column density of the cores.
Among the star-forming clouds only, we also find a tentative positive trend between \r1415\ and both 
the sonic Mach number and the dust temperature. We speculate that this can be due to the selective 
photodissociation mechanisms, which tends to decrease the \N15\ abundance, and hence to 
increase the \r1415\ ratio) in cores more exposed to UV photons. The source of these photons
is probably external for cores with higher column density, independent on the evolutionary stage
of the core. However, due to the fact that
the correlations proposed are faint, and that most \N15\ lines are detected at a significance level
of $3-5\sigma$ rms in the spectra, any conclusion that can be drawn from these trends must
be taken with caution.
\end{itemize}

{\it Acknowledgments.} 

We would like to thank the anonymous referee for their constructive feedback on the manuscript.
FF is grateful to Laura Colzi for a very useful discussion. This paper makes use of the following 
ALMA data: ADS/JAO.ALMA\#2017.1.00687.S \& ADS/JAO.ALMA\#2018.1.00850.S. ALMA is a 
partnership of ESO (representing its member states), NSF (USA) and NINS (Japan), 
together with NRC (Canada), MOST and ASIAA (Taiwan), and KASI (Republic of Korea), 
in cooperation with the Republic of Chile. The Joint ALMA Observatory is operated by 
ESO, AUI/NRAO and NAOJ. I.J.-S. has received partial support from the Spanish FEDER 
(project number ESP2017-86582-C4-1-R) and the Ministry of Science and Innovation 
through project number PID2019-105552RB-C41.
ATB would like to acknowledge funding from the European Research Council (ERC) under 
the European Union Horizon 2020 research and innovation programme (grant agreement 
No.726384/Empire).

\section*{Data availability}

Data used within this work will be shared on reasonable request to the corresponding author.

{}

\newpage

\renewcommand{\thefigure}{A-\arabic{figure}}
\setcounter{figure}{0}
\section*{Appendix A: Spectra.}
\label{appa}
We show in this appendix the spectra extracted from the contours illustrated in Fig.~\ref{fig:dendro}.

\begin{figure}
\centering
{\includegraphics[width=8.5cm,angle=0]{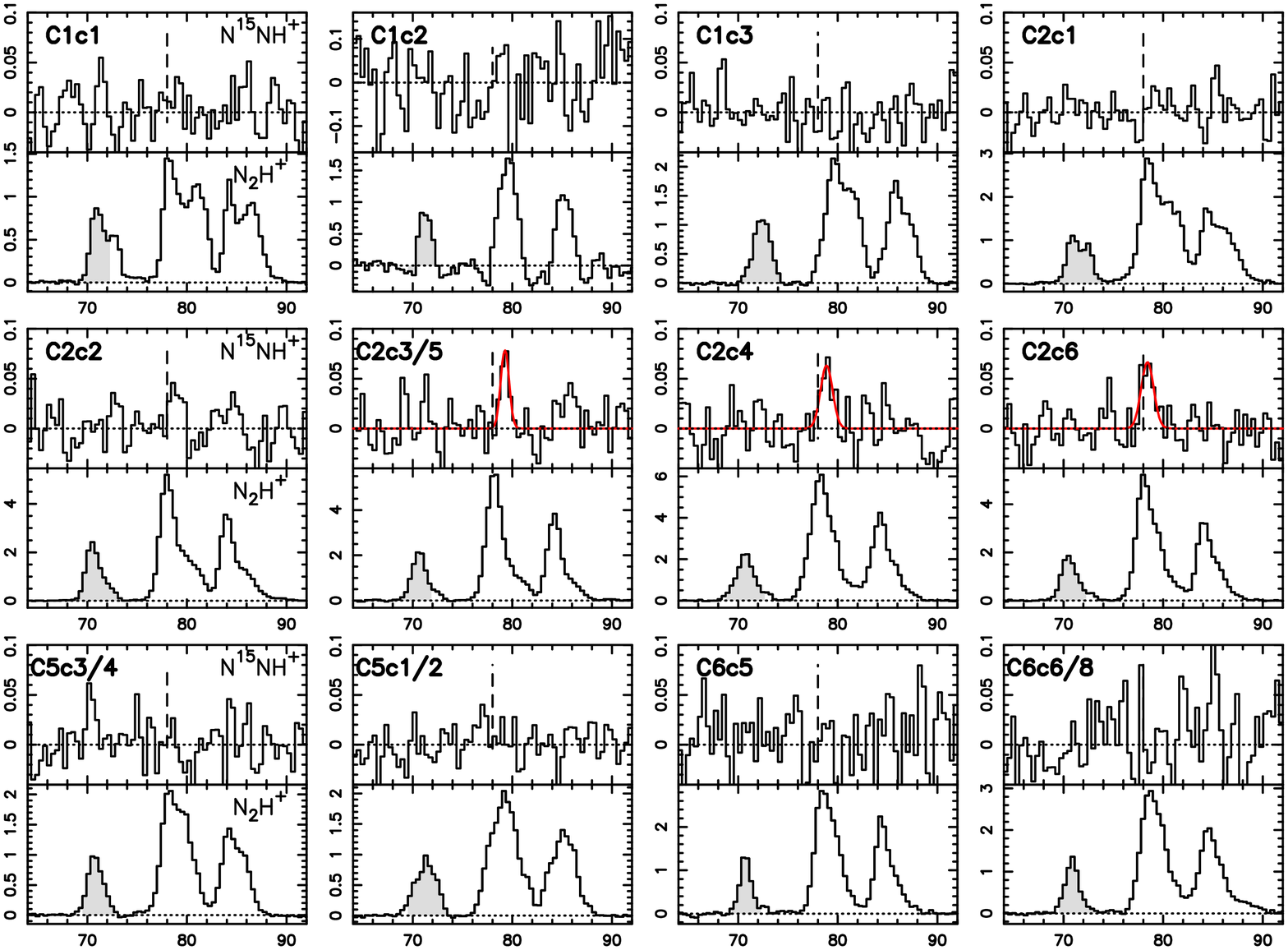}
\includegraphics[width=8.5cm,angle=0]{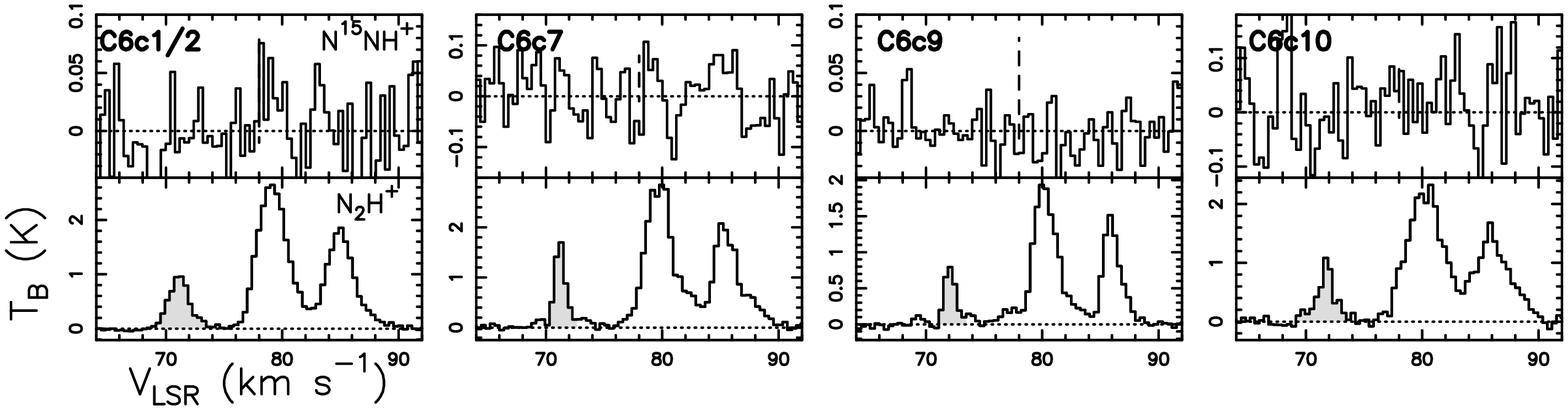}
}
\caption{Spectra of \N15\ and \H (1--0) extracted from the 3~mm continuum
cores of Cloud C identified in Table~\ref{tab:columns}. In each spectrum,
the horizontal dotted line shows the zero-rms level, and the vertical dashed line the systemic
cloud velocity (Table~\ref{tab:sources}). The red curves in the \N15\ spectra are the best fits
to the main hyperfine component towards C2c3/5, C2c4, and C2c6 because the hyperfine satellites 
are not revealed (see Sect.~\ref{n15nhp-cont}). 
The grey region in each \H\ spectrum indicates the integrated intensity used in the calculation of the
total column density.}
\label{fig:spectra-cloudC}
\end{figure}


\begin{figure}
\centering
{\includegraphics[width=8.5cm,angle=0]{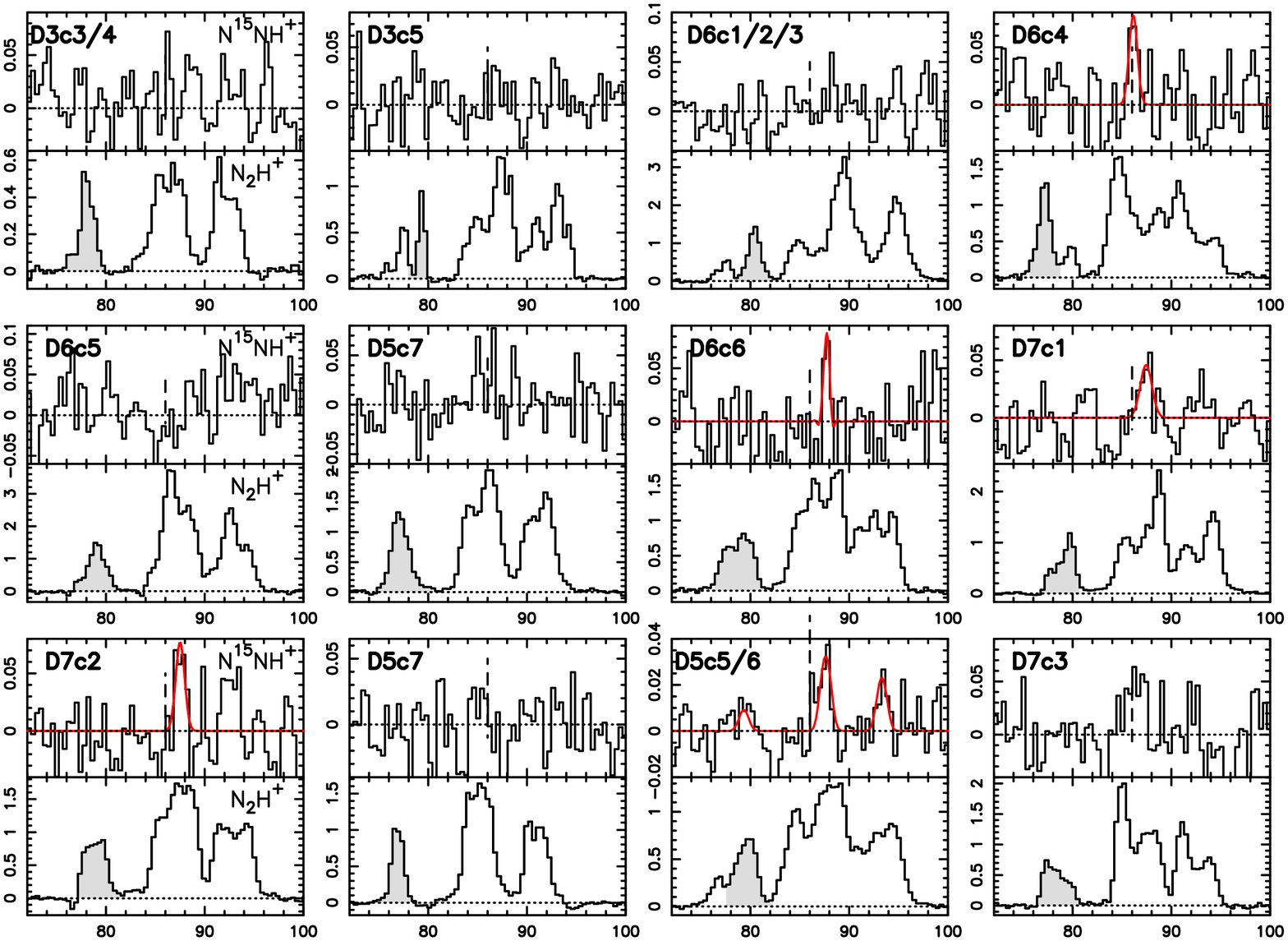}
\includegraphics[width=8.5cm,angle=0]{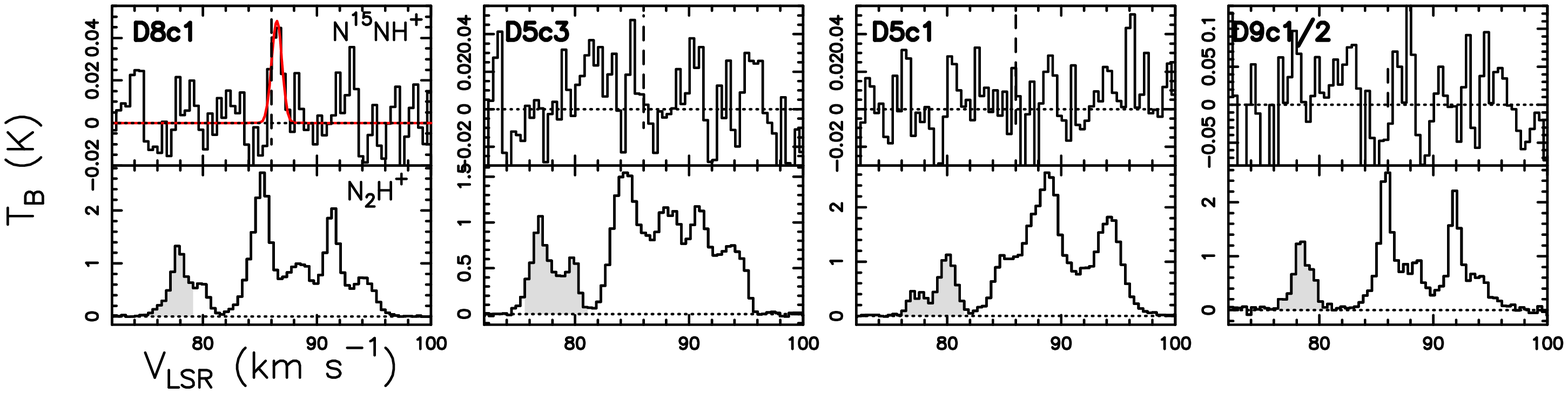}
}
\caption{Same as~\ref{fig:spectra-cloudC} for the 3~mm continuum cores in Cloud D in Table~\ref{tab:columns}.
Note that for core D5c5/6 we could fit the total hyperfine structure of the \N15(1--0) line, and not only the
main component.
}
\label{fig:spectra-cloudD}
\end{figure}


\begin{figure}
\centering
{\includegraphics[width=8.5cm,angle=0]{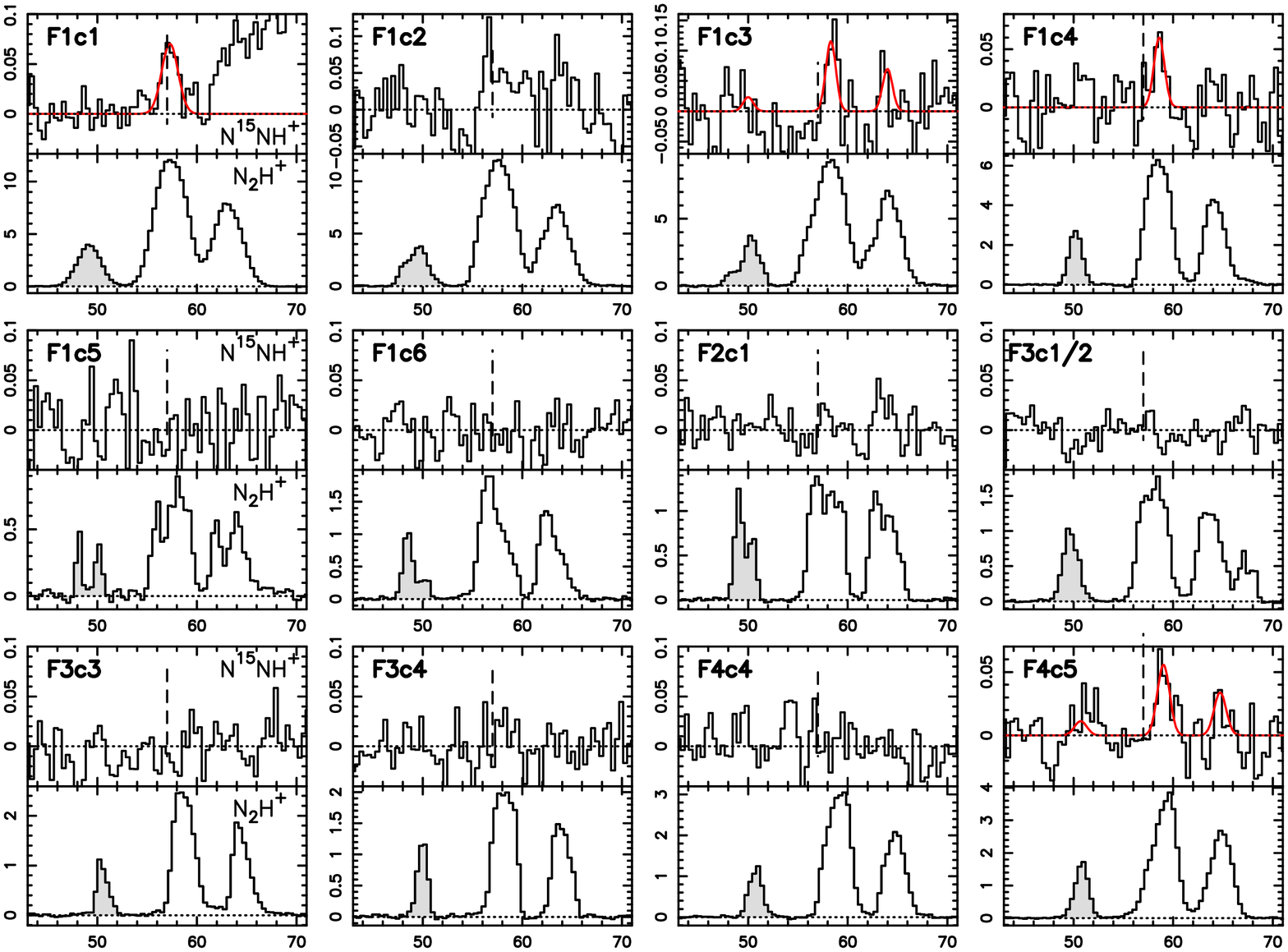}
\includegraphics[width=8.5cm,angle=0]{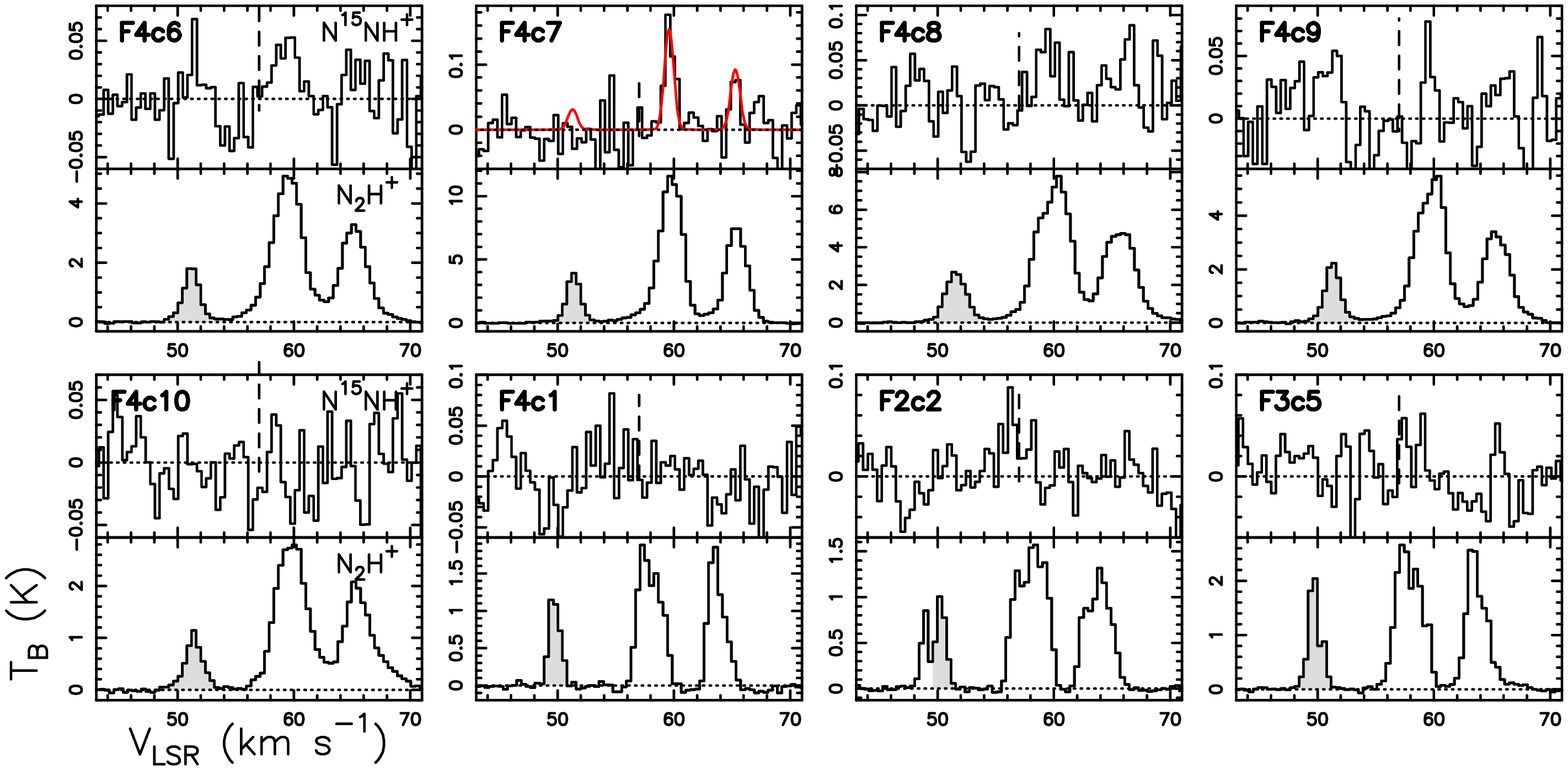}}
\caption{Same as~\ref{fig:spectra-cloudC} for the 3~mm continuum cores in Cloud F in Table~\ref{tab:columns}.
Note that we could fit the total hyperfine structure of the \N15(1--0) lines of F1c3, F4c5, and F4c7.}
\label{fig:spectra-cloudF}
\end{figure}


\begin{figure}
\centering
{\includegraphics[width=8.5cm,angle=0]{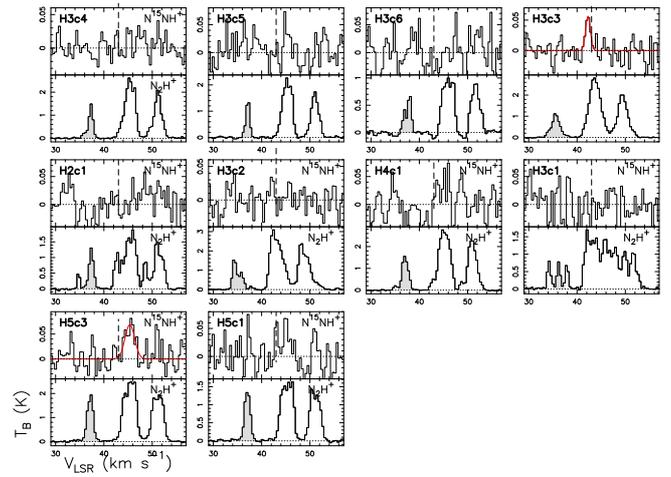}
}
\caption{Same as~\ref{fig:spectra-cloudC} for the 3~mm continuum cores of Cloud H.
}
\label{fig:spectra-cloudH}
\end{figure}

\begin{figure}
\centering
{\includegraphics[width=8.5cm,angle=0]{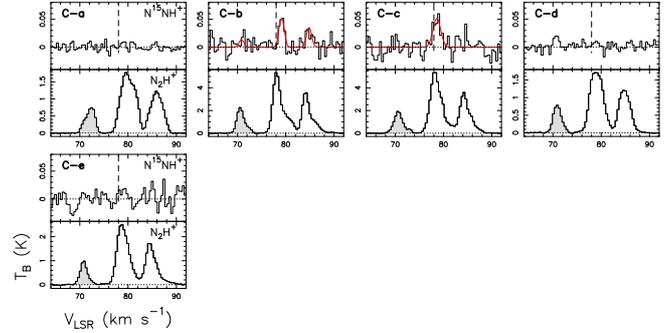}
}
\caption{Spectra of \N15\ and \H (1--0) extracted from the \H\ cores of Cloud C (Fig.~\ref{fig:dendro}) 
identified using the dendrogram analysis (paper I). All symbols, lines, and colours have the same
meaning as in Fig.~\ref{fig:spectra-cloudC}.}
\label{fig:spectra-cloudC-letters}
\end{figure}

\begin{figure}
\centering
{\includegraphics[width=8.5cm,angle=0]{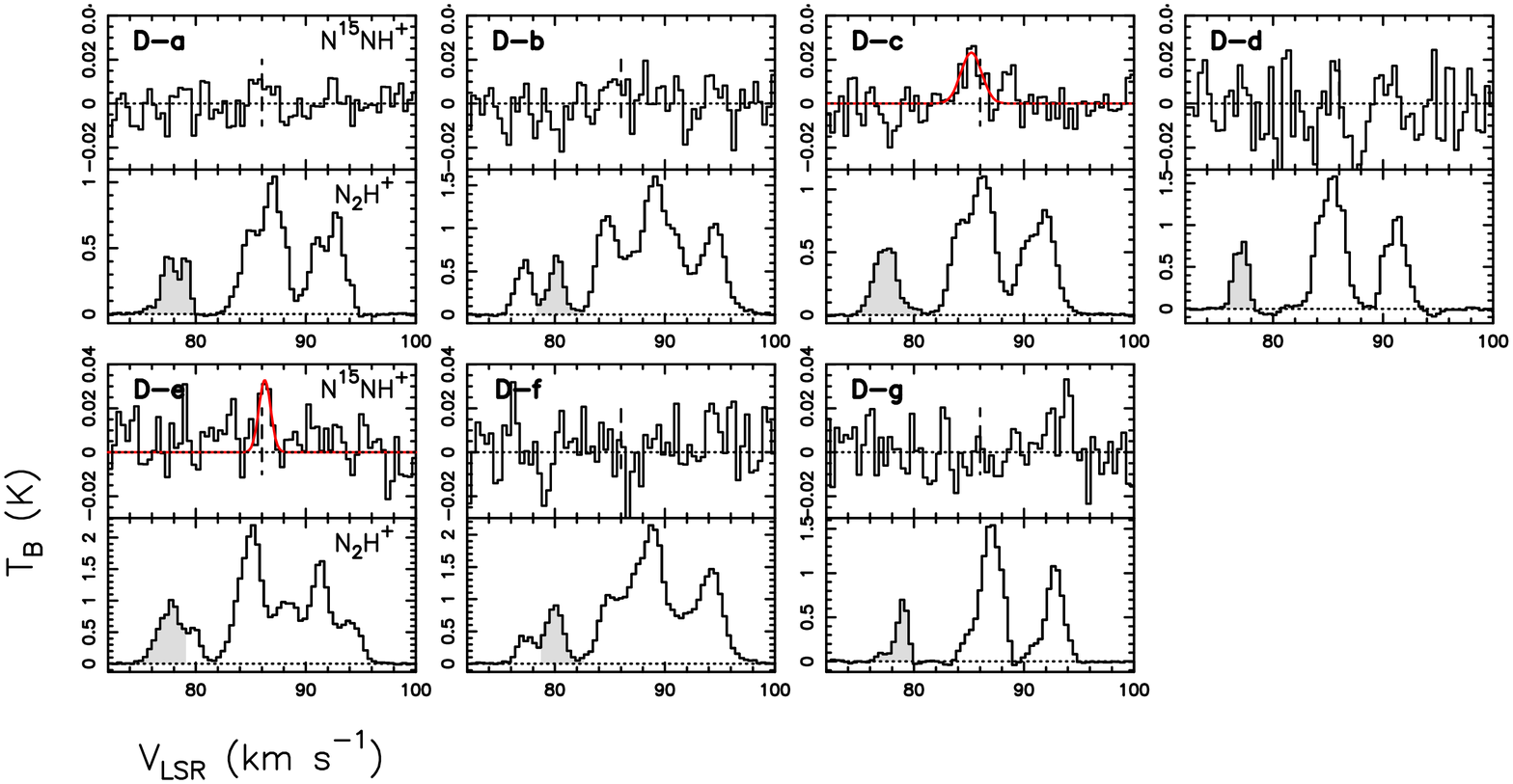}
}
\caption{Same as~\ref{fig:spectra-cloudC-letters} for Cloud D.
}
\label{fig:spectra-cloudD-letters}
\end{figure}

\begin{figure}
\centering
{\includegraphics[width=8.5cm,angle=0]{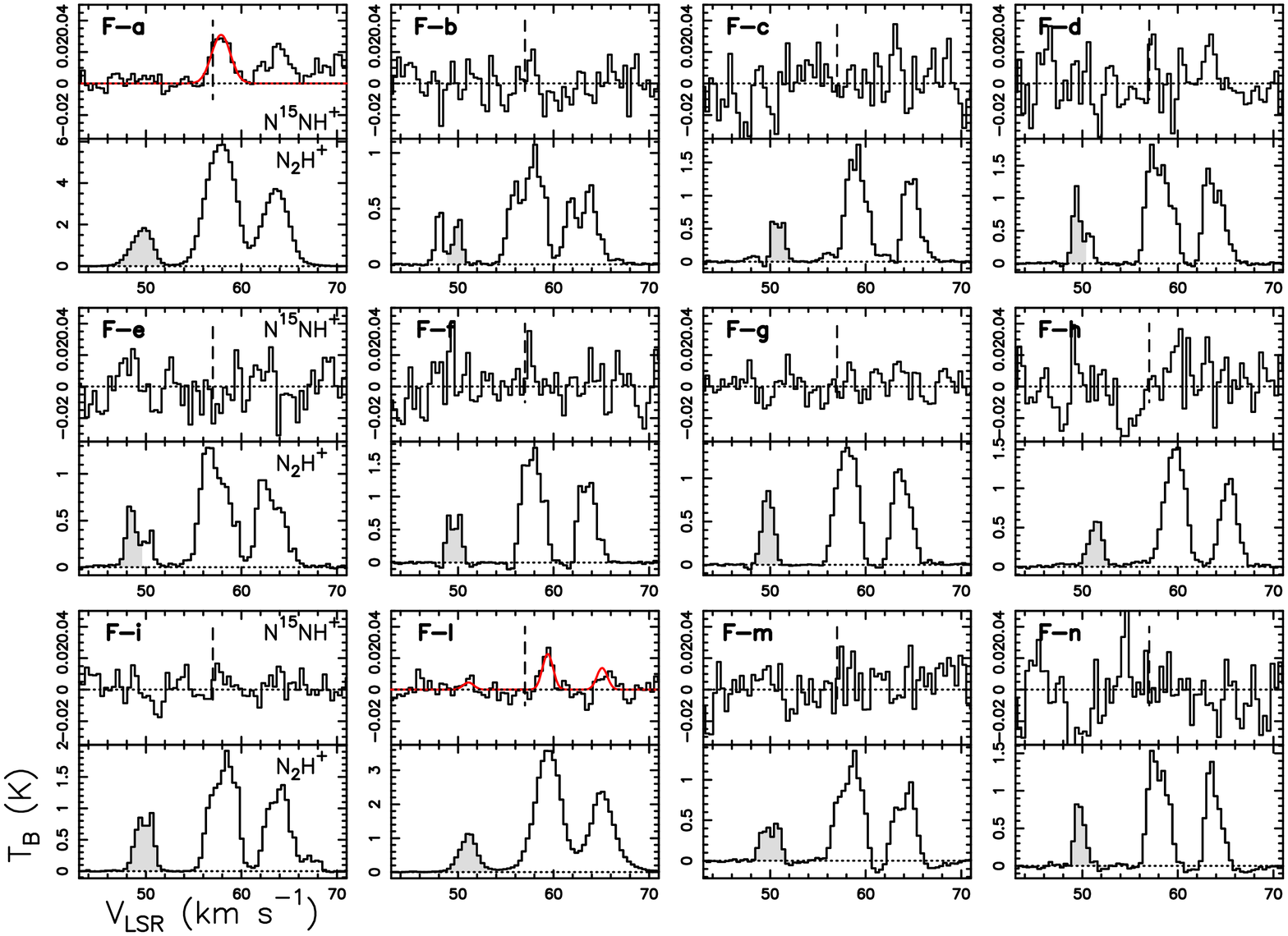}
}
\caption{Same as~\ref{fig:spectra-cloudC-letters} for Cloud F.
}
\label{fig:spectra-cloudF-letters}
\end{figure}

\begin{figure}
\centering
{\includegraphics[width=8.5cm,angle=0]{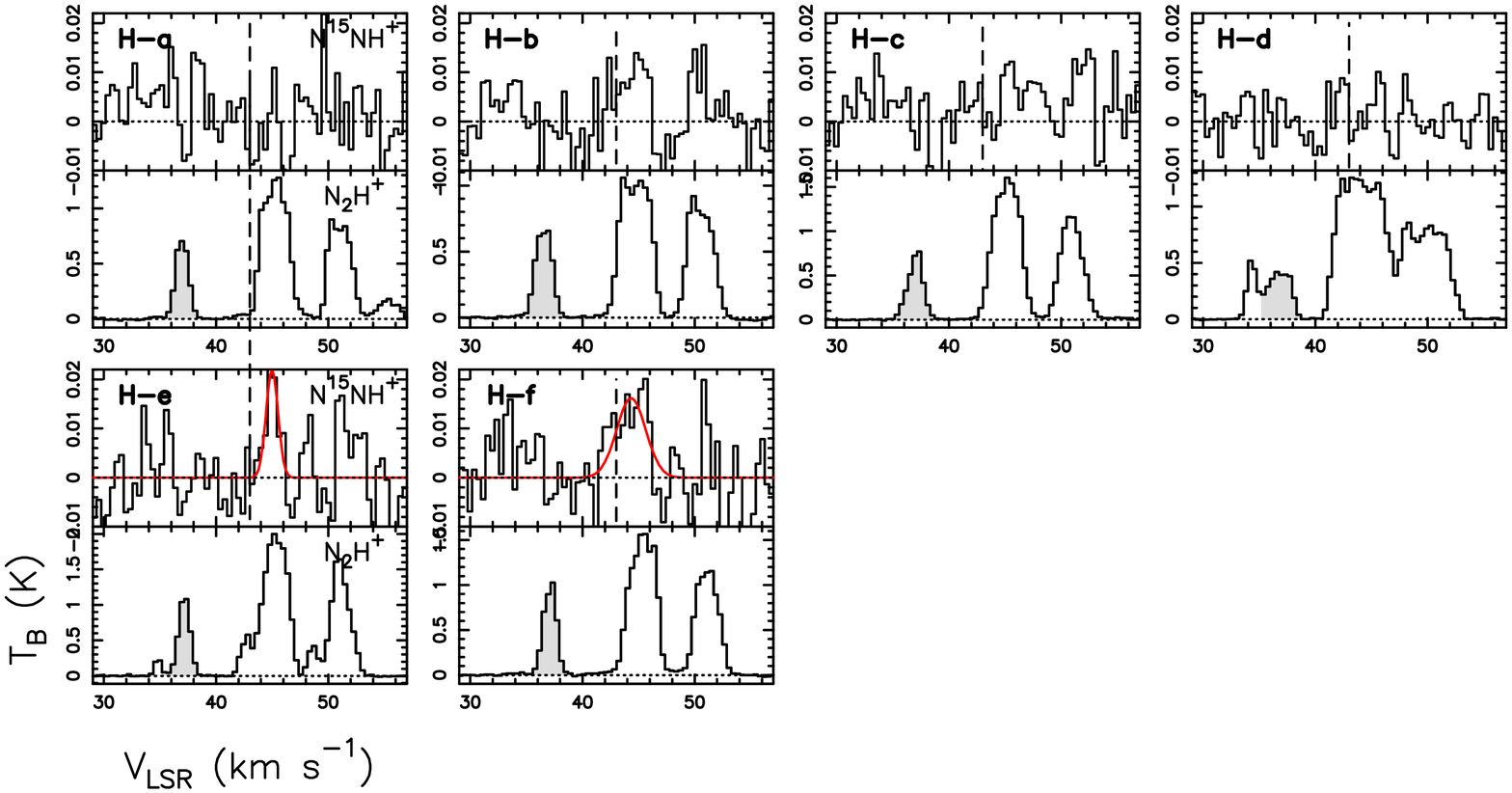}
}
\caption{Same as~\ref{fig:spectra-cloudC-letters} for Cloud H.
}
\label{fig:spectra-cloudH-letters}
\end{figure}

\end{document}